%% file: main.tex
\documentclass[twocolumn]{aastex631}
\usepackage{}

\usepackage{gensymb}  

\submitjournal{ApJ}

\shorttitle{JWST Imaging of MACS0647--JD}
\shortauthors{Hsiao et al.}

\begin{document}

\title{JWST reveals a possible $z \sim 11$ galaxy merger in triply-lensed MACS0647--JD}

\correspondingauthor{Tiger Hsiao}
\email{yhsiao17@jhu.edu}


\author[0000-0003-4512-8705]{Tiger Yu-Yang Hsiao}
\affiliation{Center for Astrophysical Sciences, Department of Physics and Astronomy, The Johns Hopkins University, 3400 N Charles St. Baltimore, MD 21218, USA}

\author[0000-0001-7410-7669]{Dan Coe}
\affiliation{Space Telescope Science Institute (STScI), 
3700 San Martin Drive, Baltimore, MD 21218, USA}
\affiliation{Association of Universities for Research in Astronomy (AURA), Inc.
for the European Space Agency (ESA)}
\affiliation{Center for Astrophysical Sciences, Department of Physics and Astronomy, The Johns Hopkins University, 3400 N Charles St. Baltimore, MD 21218, USA}

\author[0000-0002-5258-8761]{Abdurro'uf}
\affiliation{Center for Astrophysical Sciences, Department of Physics and Astronomy, The Johns Hopkins University, 3400 N Charles St. Baltimore, MD 21218, USA}
\affiliation{Space Telescope Science Institute (STScI), 
3700 San Martin Drive, Baltimore, MD 21218, USA}

\author[0000-0003-1432-7744]{Lily Whitler}
\affiliation{Steward Observatory, University of Arizona, 933 N Cherry Ave, Tucson, AZ 85721 USA}

\author[0000-0003-1187-4240]{Intae Jung}
\affiliation{Space Telescope Science Institute (STScI), 3700 San Martin Drive, Baltimore, MD 21218, USA}

\author[0000-0002-3475-7648]{Gourav Khullar}
\affiliation{Department of Physics and Astronomy and PITT PACC, University of Pittsburgh, Pittsburgh, PA 15260, USA}

\author[0000-0002-7876-4321]{Ashish Kumar Meena}
\affiliation{Physics Department, Ben-Gurion University of the Negev, P.O. Box 653, Be'er-Sheva 84105, Israel}

\author[0000-0001-8460-1564]{Pratika Dayal}
\affiliation{Kapteyn Astronomical Institute, University of Groningen, P.O. Box 800, 9700 AV Groningen, The Netherlands}

\author[0000-0002-8638-1697]{Kirk S.~S.~Barrow}
\affiliation{Department of Astronomy, University of Illinois at Urbana-Champaign, 
1002 W Green St, Urbana, IL 61801, USA}

\author[0000-0002-8763-3702]{Lillian Santos-Olmsted}
\affiliation{Kavli Institute for Particle Astrophysics and Cosmology, Stanford University, 452 Lomita Mall, Stanford, CA 94305-4085, USA}

\author[0000-0001-9055-7045]{Adam Casselman}
\affiliation{Department of Aerospace Engineering, University of Illinois at Urbana, Champaign 104 S Wright St, Urbana, IL 61801, USA}

\author[0000-0002-5057-135X]{Eros Vanzella}
\affiliation{INAF -- OAS, Osservatorio di Astrofisica e Scienza dello Spazio di Bologna, via Gobetti 93/3, I-40129 Bologna, Italy}

\author[0000-0001-6342-9662]{Mario Nonino}
\affiliation{INAF-Trieste Astronomical Observatory, Via Bazzoni 2, 34124, Trieste, Italy}

\author[0000-0002-6090-2853]{Yolanda Jim\'enez-Teja}
\affiliation{Instituto de Astrof\'isica de Andaluc\'ia, Glorieta de la Astronom\'ia s/n, 18008 Granada, Spain}
\affiliation{Observatório Nacional - MCTI (ON), Rua Gal. José Cristino 77, São Cristóvão, 20921-400, Rio de Janeiro, Brazil}

\author[0000-0003-3484-399X]{Masamune Oguri}
\affiliation{Center for Frontier Science, Chiba University, 1-33 Yayoi-cho, Inage-ku, Chiba 263-8522, Japan}
\affiliation{Department of Physics, Graduate School of Science, Chiba University, 1-33 Yayoi-Cho, Inage-Ku, Chiba 263-8522, Japan}

\author{Daniel P. Stark}
\affiliation{Steward Observatory, University of Arizona, 933 N Cherry Ave, Tucson, AZ 85721 USA}

\author[0000-0001-6278-032X]{Lukas J. Furtak}
\affiliation{Physics Department, Ben-Gurion University of the Negev, P.O. Box 653, Be'er-Sheva 84105, Israel}

\author[0000-0002-0350-4488]{Adi Zitrin}
\affiliation{Physics Department, Ben-Gurion University of the Negev, P.O. Box 653, Be'er-Sheva 84105, Israel}

\author[0000-0002-0786-7307]{Angela Adamo}
\affiliation{Department of Astronomy, Oskar Klein Centre, Stockholm University, AlbaNova University Centre, SE-106 91 Stockholm, Sweden}

\author[0000-0003-2680-005X]{Gabriel Brammer}
\affiliation{Cosmic Dawn Center (DAWN), Copenhagen, Denmark}
\affiliation{Niels Bohr Institute, University of Copenhagen, Jagtvej 128, Copenhagen, Denmark}

\author[0000-0002-7908-9284]{Larry Bradley}
\affiliation{Space Telescope Science Institute (STScI), 3700 San Martin Drive, Baltimore, MD 21218, USA}

\author[0000-0001-9065-3926]{Jose M. Diego}
\affiliation{Instituto de F\'isica de Cantabria (CSIC-UC). Avda. Los Castros s/n. 39005 Santander, Spain}

\author[0000-0003-1096-2636]{Erik Zackrisson}
\affiliation{Observational Astrophysics, Department of Physics and Astronomy, Uppsala University, Box 516, SE-751 20 Uppsala, Sweden}

\author[0000-0001-8519-1130]{Steven L. Finkelstein}
\affiliation{Department of Astronomy, The University of Texas at Austin, Austin, TX 78712}

\author[0000-0001-8156-6281]{Rogier A. Windhorst} 
\affiliation{School of Earth and Space Exploration, Arizona State University, Tempe, AZ 85287-1404, USA}

\author[0000-0003-0883-2226]{Rachana Bhatawdekar}
\affiliation{European Space Agency, ESA/ESTEC, Keplerlaan 1, 2201 AZ Noordwijk, NL}

\author[0000-0001-6251-4988]{Taylor A. Hutchison}
\altaffiliation{NASA Postdoctoral Fellow}
\affiliation{Observational Cosmology Lab, NASA Goddard Space Flight Center, Greenbelt, MD 20771, USA}

\author[0000-0002-8785-8979]{Tom Broadhurst}
\affiliation{Department of Theoretical Physics, University of the Basque Country UPV/EHU, Bilbao, Spain}
\affiliation{Donostia International Physics Center (DIPC), 20018 Donostia, Spain}
\affiliation{IKERBASQUE, Basque Foundation for Science, Bilbao, Spain}

\author[0000-0001-7399-2854]{Paola Dimauro}
\affiliation{INAF - Osservatorio Astronomico di Roma, via di Frascati 33, 00078 Monte Porzio Catone, Italy}

\author[0000-0002-8144-9285]{Felipe Andrade-Santos}
\affiliation{Department of Liberal Arts and Sciences, Berklee College of Music, 7 Haviland Street, Boston, MA 02215, USA}
\affiliation{Center for Astrophysics {$\|$} Harvard {$\&$} Smithsonian, 60 Garden Street, Cambridge, MA 02138, USA}

\author[0000-0002-1722-6343]{Jan J. Eldridge}
\affiliation{Department of Physics, University of Auckland, Private Bag 92019, Auckland, New Zealand}


\author[0000-0003-3108-9039]{Ana Acebron}
\affiliation{Dipartimento di Fisica, Universit\`a degli Studi di Milano, Via Celoria 16, I-20133 Milano, Italy}
\affiliation{INAF - IASF Milano, via A. Corti 12, I-20133 Milano, Italy}

\author[0000-0001-9364-5577]{Roberto J. Avila}
\affiliation{Space Telescope Science Institute (STScI),  3700 San Martin Drive,  Baltimore, MD 21218, USA}

\author[0000-0003-1074-4807]{Matthew B. Bayliss}
\affiliation{Department of Physics, University of Cincinnati, Cincinnati, OH 45221, USA}

\author{Alex Ben\'itez}
\affiliation{King's College London, University of London, Strand, London WC2R 2LS, UK}

\author[0000-0002-0808-4136]{Christian Binggeli}
\affiliation{Observational Astrophysics, Department of Physics and Astronomy, Uppsala University, Box 516, SE-751 20 Uppsala, Sweden}

\author[0000-0002-7365-4131]{Patricia Bolan}
\affiliation{Department of Physics and Astronomy, University of California, Davis, 1 Shields Ave, Davis, CA 95616, USA}

\author[0000-0001-5984-0395]{Maru\v{s}a Brada\v{c}}
\affiliation{Department of Mathematics and Physics, University of Ljubljana, Jadranska ulica 19, SI-1000 Ljubljana, Slovenia}
\affiliation{Department of Physics and Astronomy, University of California, Davis, 1 Shields Ave, Davis, CA 95616, USA}

\author[0000-0002-1482-5818]{Adam C. Carnall}
\affiliation{Institute for Astronomy, University of Edinburgh, Royal Observatory, Edinburgh EH9 3HJ, UK}

\author[0000-0003-1949-7638]{Christopher J. Conselice}
\affiliation{Jodrell Bank Centre for Astrophysics, University of Manchester, Oxford Road, Manchester UK}

\author[0000-0002-2808-0853]{Megan Donahue}
\affiliation{Michigan State University, Physics \& Astronomy Department, East Lansing, MI, USA}

\author[0000-0003-1625-8009]{Brenda Frye}
\affiliation{Department of Astronomy, Steward Observatory, University of Arizona, 933 North Cherry Avenue, Tucson, AZ 85721, USA}

\author[0000-0001-7201-5066]{Seiji Fujimoto}\altaffiliation{Hubble Fellow}
\affiliation{Department of Astronomy, The University of Texas at Austin, Austin, TX 78712}

\affiliation{Cosmic Dawn Center (DAWN), Copenhagen, Denmark}
\affiliation{Niels Bohr Institute, University of Copenhagen, Jagtvej 128, Copenhagen, Denmark}

\author[0000-0002-6586-4446]{Alaina Henry}
\affiliation{Space Telescope Science Institute (STScI), 
3700 San Martin Drive, 
Baltimore, MD 21218, USA}
\affiliation{Center for Astrophysical Sciences, Department of Physics and Astronomy, The Johns Hopkins University, 3400 N Charles St. Baltimore, MD 21218, USA}

\author[0000-0003-4372-2006]{Bethan L. James}
\affiliation{Space Telescope Science Institute (STScI), 
3700 San Martin Drive, Baltimore, MD 21218, USA}

\author[0000-0002-3838-8093]{Susan A. Kassin}
\affiliation{Space Telescope Science Institute (STScI), 
3700 San Martin Drive, Baltimore, MD 21218, USA}
\affiliation{Center for Astrophysical Sciences, Department of Physics and Astronomy, The Johns Hopkins University, 3400 N Charles St. Baltimore, MD 21218, USA}

\author[0000-0001-8152-3943]{Lisa Kewley}
\affiliation{Center for Astrophysics {$\|$} Harvard {$\&$} Smithsonian, 60 Garden Street, Cambridge, MA 02138, USA}

\author[0000-0003-2366-8858]{Rebecca L. Larson}
\altaffiliation{NSF Graduate Fellow}
\affiliation{Department of Astronomy, The University of Texas at Austin, Austin, TX 78712}

\author[0000-0003-3234-7247]{Tod Lauer}
\affiliation{NSF’s National Optical-Infrared Astronomy Research Laboratory, 950 North Cherry Avenue, Tucson, AZ 85719, USA}

\author[0000-0002-9402-186X]{David Law}
\affiliation{Space Telescope Science Institute (STScI), 
3700 San Martin Drive, 
Baltimore, MD 21218, USA}

\author[0000-0003-3266-2001]{Guillaume Mahler}
\affiliation{Institute for Computational Cosmology, Durham University, South Road, Durham DH1 3LE, UK}
\affiliation{Centre for Extragalactic Astronomy, Durham University, South Road, Durham DH1 3LE, UK}

\author[0000-0003-0094-6827]{Ramesh Mainali}
\affiliation{Observational Cosmology Lab, NASA Goddard Space Flight Center, Greenbelt, MD 20771, USA}

\author[0000-0003-0503-4667]{Stephan McCandliss}
\affiliation{Center for Astrophysical Sciences, Department of Physics and Astronomy, The Johns Hopkins University, 
3400 N Charles St. 
Baltimore, MD 21218, USA}

\author[0000-0003-0892-5203]{David Nicholls}
\affiliation{Research School of Astronomy and Astrophysics, Australian National University, Canberra, ACT, Australia}
\affiliation{ARC Centre of Excellence for All Sky Astrophysics in 3 Dimensions (ASTRO 3D), Australia}

\author[0000-0003-3382-5941]{Norbert Pirzkal}
\affiliation{Space Telescope Science Institute (STScI), 
3700 San Martin Drive, 
Baltimore, MD 21218, USA}

\author[0000-0002-9365-7989]{Marc Postman}
\affiliation{Space Telescope Science Institute (STScI), 
3700 San Martin Drive, 
Baltimore, MD 21218, USA}

\author[0000-0002-7627-6551]{Jane R.~Rigby}
\affiliation{Observational Cosmology Lab, NASA Goddard Space Flight Center, Greenbelt, MD 20771, USA}

\author[0000-0003-0894-1588]{Russell Ryan}
\affiliation{Space Telescope Science Institute (STScI), 
3700 San Martin Drive, 
Baltimore, MD 21218, USA}

\author[0000-0002-9132-6561]{Peter Senchyna}
\affiliation{Carnegie Observatories, 813 Santa Barbara Street, Pasadena, CA 91101, USA}

\author[0000-0002-7559-0864]{Keren Sharon}
\affiliation{Department of Astronomy, University of Michigan, 1085 S. University Ave, Ann Arbor, MI 48109, USA}

\author{Ikko Shimizu}
\affiliation{Department of Literature, Shikoku Gakuin University, 3-2-1 Bunkyocho, Zentsuji, Kagawa 765-8505, Japan}

\author[0000-0002-6338-7295]{Victoria Strait}
\affiliation{Cosmic Dawn Center (DAWN), Copenhagen, Denmark}
\affiliation{Niels Bohr Institute, University of Copenhagen, Jagtvej 128, Copenhagen, Denmark}

\author[0000-0001-5940-338X]{Mengtao Tang}
\affiliation{Department of Physics and Astronomy, University College London, Gower Street, London WC1E 6BT, UK}

\author[0000-0001-9391-305X]{Michele Trenti}
\affiliation{School of Physics, University of Melbourne, Parkville VIC 3010, Australia}
\affiliation{ARC Centre of Excellence for All-Sky Astrophysics in 3 Dimensions, University of Melbourne, Parkville VIC 3010, Australia}

\author[0000-0002-4853-1076]{Anton Vikaeus}
\affiliation{Observational Astrophysics, Department of Physics and Astronomy, Uppsala University, Box 516, SE-751 20 Uppsala, Sweden}

\author[0000-0003-1815-0114]{Brian Welch}
\affiliation{Department of Astronomy, University of Maryland, College Park, MD 20742, USA}
\affiliation{Observational Cosmology Lab, NASA Goddard Space Flight Center, Greenbelt, MD 20771, USA}
\affiliation{Center for Research and Exploration in Space Science and Technology, NASA/GSFC, Greenbelt, MD 20771}


\input{newcommands}

\begin{abstract}
MACS0647--JD is a triply-lensed $z \sim 11$ galaxy originally discovered with the \emph{Hubble Space Telescope}.
The three lensed images are magnified by factors of \about8, 5, and 2 to AB mag 25.1, 25.6, and 26.6 at 3.5 \um.
{The brightest is over a magnitude brighter than other galaxies recently discovered at similar redshifts $z > 10$ with \JWST.}
Here we report new \JWST\ imaging clearly resolves MACS0647--JD as having two components
that are either merging galaxies or stellar complexes within a single galaxy.
The brighter larger component ``A'' is intrinsically very blue {($\beta \sim -2.6\pm0.1$)},
likely due to very recent star formation and no dust, 
and is spatially extended with an effective radius 
$\sim70${$\pm 24$} {\rm pc}.
The smaller component ``B'' ($r \sim20${$^{+8}_{-5}\,$}${\rm pc}$)
appears redder {($\beta \sim -2\pm0.2$)}, 
likely because it is older (100--200\,Myr)
with mild dust extinction ($A_V \sim 0.1$\,mag).
With an estimated stellar mass ratio of roughly 2:1
and physical projected separation $\sim400\,{\rm pc}$, 
we may be witnessing a galaxy merger 430 million years after the Big Bang.
We identify galaxies with similar colors in a high-redshift simulation,
finding their star formation histories to be {dissimilar, which is also suggested by the SED fitting,}
suggesting they formed further apart.
We also identify a candidate companion galaxy ``C'' $\sim$3\,kpc away, likely destined to merge with A and B.
Upcoming \JWST\ NIRSpec observations planned for January 2023
will deliver spectroscopic redshifts and more physical properties
for these tiny magnified distant galaxies observed in the early universe.

\end{abstract}


\section{Introduction}
\label{sec:intro}

{
Galaxies have formed from the repeated mergers of small star-forming clumps over cosmic time, with some small galaxies left over even today, such as the Magellanic clouds.
\JWST\ has now discovered two such small galaxies within the first 430 million years, that are seen close to the very start of this process.
Studies have shown that {up to $85\%$ of 
present-day massive 
galaxies} went through a galaxy merger in their lifetime, indicating that galaxy mergers play an important role in the formation and evolution of galaxies {\citep[e.g.,][]{Bell2006,Stewart2009,Hopkins2010,Lotz2011,Rodriguez2015,Duncan2019,Sotillo2022}}. 
The Milky Way itself likely experienced a major merger at $z\sim2$ with the so-called Gaia-Sausage-Enceladus galaxy {\citep{Helmi2018,Belokurov2018,Bonaca2020,Naidu2021,Xiang2022}}. 
Based on reconstructions of this event, \citet{Naidu2021} concluded that $\approx 50\%$ of the stellar mass of the current halo of the Milky Way came from this galaxy.
More generally speaking, mergers build up the stellar content and transform galaxy morphology \citep[e.g.,][]{Toomre1972,Barnes1992,Mihos1996,Husko2023}
Mergers are also believed to affect the kinematics and distribution of stars \citep[e.g.,][]{Naab2009,Dokkum2010,Newman2012}, and play a key role in the growth of supermassive black holes \citep[e.g.,][]{Treister2012,Ellison2019,Zhang2023}.}

{\JWST} (\citealt{Gardner06_jwst}) is a state-of-the-art infrared space-based telescope which was launched in December 2021 and started scientific observations recently in July 2022 \citep{Rigby22_jwst}.
Numerous high-redshift candidates have been discovered based on their photometric redshifts and drop-out selections 
\citep[e.g.,][]{Naidu2022,Castellano2022,Adams2022,Yan2022, Donnan2022, Harikane2022,Atek2022,Finkelstein2022,Whitler2022b,Bradley2022,Castellano2022}.
Within its first few months, {\JWST} is quickly transforming our understanding of the early Universe 
(e.g., with flat / disky galaxies reported at $z \sim 2$ -- 6:
\citealt{Ferreira22,Nelson2022}).


Gravitational lensing by massive galaxy clusters magnifies the light and sizes of distant objects.
Thanks to these cosmic telescopes, not only are the fluxes of faint objects in the early Universe are boosted to the observable regime, the sizes of small-scale structures are amplified  \citep[e.g.,][]{Welch22_clumps,Vanzella2022,Mestric2022,Welch2022JWST, claeyssens2022}.
Thus, lensing has enabled us to discover early galaxies and study their properties \citep[e.g,][]{Coe2013}.
In order to study several key scientific topics in the early Universe, several lensing cluster surveys have been conducted, including the Cluster Lensing and Supernova survey with Hubble \citep[CLASH\footnote{\url{https://www.stsci.edu/~postman/CLASH/}};][]{Postman2012}, the Hubble Frontier Fields \citep{Lotz2017}, and the Reionization Lensing Cluster Survey \citep[RELICS\footnote{\url{https://relics.stsci.edu}}][]{Coe2019}.

CLASH is one of the large Hubble treasury programs which adopted the lensing technique to study distant galaxies \citep[e.g.,][]{Zheng2012,Coe2013,Bouwens2014,Smit2014,Bradley2014}, supernovae and cosmology \citep[e.g.,][]{Graur2014,Patel2014,Rodney2014,Strolger2015,Riess2018,Gomez2018}, dark matter in clusters \citep[e.g.,][]{Pacucci2013,Eichner2013,Sartoris2014,Umetsu2014,Merten2015}, and galaxies in clusters \citep[e.g.,][]{Postman2012b,Burke2015,Donahue2015,Fogarty2017,Connor2017}.

CLASH imaged 25 massive galaxy clusters in 16 filters with {\HST} from near-UV ($\sim200\,{\rm nm}$) to near-IR ($\sim1.6\,{\rm \mu m}$).
Five of the clusters were selected for their strong lensing strength,
including MACSJ0647.7+7015 \citep[MACS0647; $z=0.591$;][]{Ebeling2007}
modeled by \cite{Zitrin2011}.
CLASH observations of MACSJ0647.7+7015 revealed 32 lensed $z \sim 6$ -- 8 candidates \citep{Bradley2014}
and a triply-lensed $z\sim11$ candidate MACS0647--JD \citep{Coe2013}.

MACS0647--JD had a photometric redshift $z = 10.7^{+0.6}_{-0.4}$
based on \HST\ images,
where it was detected in only the two reddest filters F140W and F160W,
dropping out of 15 bluer filters, including $J$ F125W, 
hence the name JD (J-band dropout).
Despite lensing magnifications up to a factor of 8,
MACS0647--JD was spatially unresolved in \HST\ imaging.

\begin{figure}
\includegraphics[width=\columnwidth]{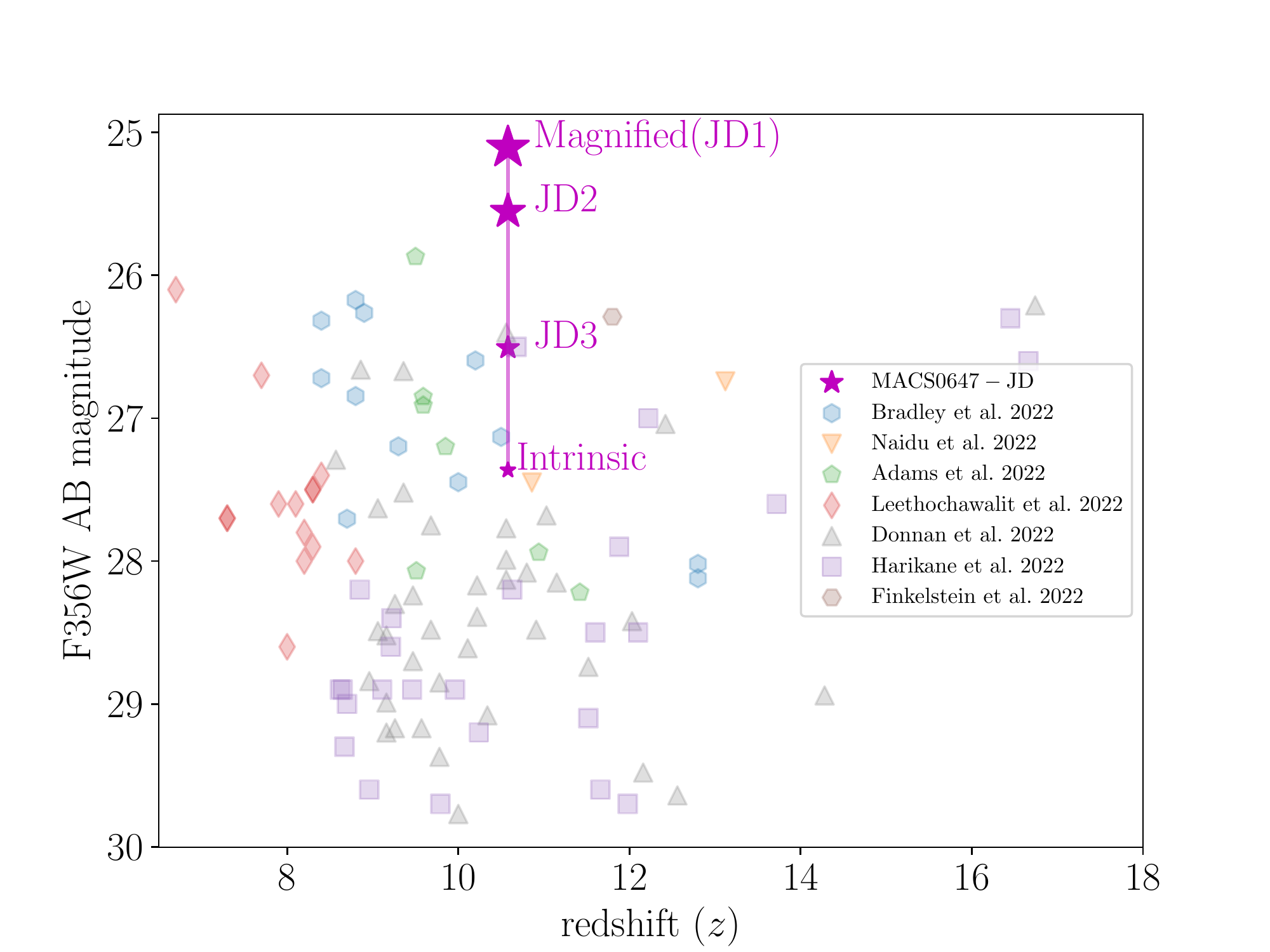}
\caption{F356W AB magnitude vs.~redshift for MACS0647--JD
and $z\sim 8$ -- 16 candidates recently discovered in \JWST\ imaging
\citep{Bradley2022,Naidu2022,Adams2022,Leethochawalit2022,Donnan2022,Harikane2022,Finkelstein2022}.
MACS0647--JD is shown both as observed (magnified) and delensed (intrinsic) according to our models.
\label{fig:abmag}
}
\end{figure}

MACS0647--JD was the first robust $z\sim11$ candidate of the \HST\ era,
followed by GN-z11, which surpassed it at $z=11.1$ \citep{Oesch16}
and is similarly bright (F160W AB mag $\sim$26)
without the benefit of lensing magnification.
{Lensed as brightly as F356W AB mag 25.1,
MACS0647--JD is 1 -- 5 magnitudes brighter than recently discovered $z \sim 8$ -- 16 candidates
reported in \JWST\ imaging
\citep[Figure \ref{fig:abmag};][]{Bradley2022,Naidu2022,Adams2022,Atek2022,Leethochawalit2022,Donnan2022,Harikane2022,Finkelstein2022}.
Because it is so bright, we can study its physical properties in more detail.
More detail about the photometry measurement and the lensing will be later described in Section \ref{sec:data} and \ref{sec:lens}}.


\cite{Pirzkal15} analyzed \HST\ WFC3/IR G141 grism spectroscopy
of MACS0647--JD,
concluding that any emission line bright enough to reproduce the observed photometry was ruled out by the observations. 
Any line or combination of emission lines with a flux of 
$10^{-17}$\,erg/s/cm$^2$/\AA\ would have been detected at the 5$\sigma$\ level,
adding further support for $z \sim 11$
excluding a lower redshift interloper, as in \cite{Brammer13}.

Lens modeling contributes geometric redshift corroboration
based on the measured separations between the lensed images.
The models in \citet{Coe2013} and \cite{Chan17}
both supported $z \sim 11$.

\citet{Lam2019} analyzed deep {\it Spitzer} imaging (50 hours / band),
modeling and subtracting light of nearby galaxies
to arrive at tentative detections of MACS0647--JD.
Photometry varied between the three lensed images,
yielding estimates of 
stellar mass $M_{*}/M_{\odot} \sim 10^8$ -- $10^9$,
specific star formation rates $\sim$~3 -- 10\,Gyr$^{-1}$,
and ages ranging between $\sim$~10\,Myr -- 400\,Myr
(the age of the universe at $z \sim 11$).



\JWST\ observing program GO 1433 (PI Coe) aims at studying MACS0647--JD in more detail, 
obtaining higher resolution images, measuring colors, 
and obtaining spectroscopy to more precisely measure the redshift
and constrain other physical properties including metallicity.
NIRCam imaging was obtained in 6 filters spanning 1.0--5.0\,\um,
out to $\sim$4300\,\AA\ rest-frame at $z=10.6$.
The second epoch of observations planned for January 2023
will obtain NIRSpec MSA PRISM observations
and add the NIRCam F480M filter, fully redward of the Balmer break
to obtain better measurements of ages and stellar masses at $z\sim11$.
All data from this program are public.
We are releasing high-level science products and analysis tools 
online.\footnote{\url{https://cosmic-spring.github.io}} 

In this paper, we report new observations of MACS0647--JD with 6 \JWST\ NIRCam filters
and derive physical properties including the stellar mass and dust content,
while constraining the star formation history.
This paper is organized as follows.
In \S\ref{sec:data}, we describe the \JWST\ and \HST\ observational data and the data reduction.
In \S\ref{sec:galaxies}, we present the detected objects and their sizes and separations based on lens modeling.
We detail photometry measurements in \S\ref{sec:photometry} and SED fitting in \S\ref{sec:sed}.
In \S\ref{sec:results&discussion}, we discuss our results, including measurements of physical parameters from SED fitting.
We also present properties of analog galaxies identified with similar colors in a hydrodynamic simulation.
We summarize our conclusions in \S\ref{sec:conclusions}.

We adopt the AB magnitude system
\citep[$m_{\rm AB} = 31.4 - 2.5 \log(f_\nu / {\rm nJy})$;][]{Oke74,OkeGunn83}
%
and the {\em Planck} 2018 flat \LCDM\ cosmology \citep{Planck18_cosmo}
with $H_0 = 67.7$ km s\inv\ Mpc\inv, $\Om = 0.31$, and $\OL = 0.69$,
for which the universe is 13.8 billion years old,
and $1 \arcsec \sim 4$ kpc at $z \sim 11$.

\begin{figure*}
\includegraphics[width=\textwidth]{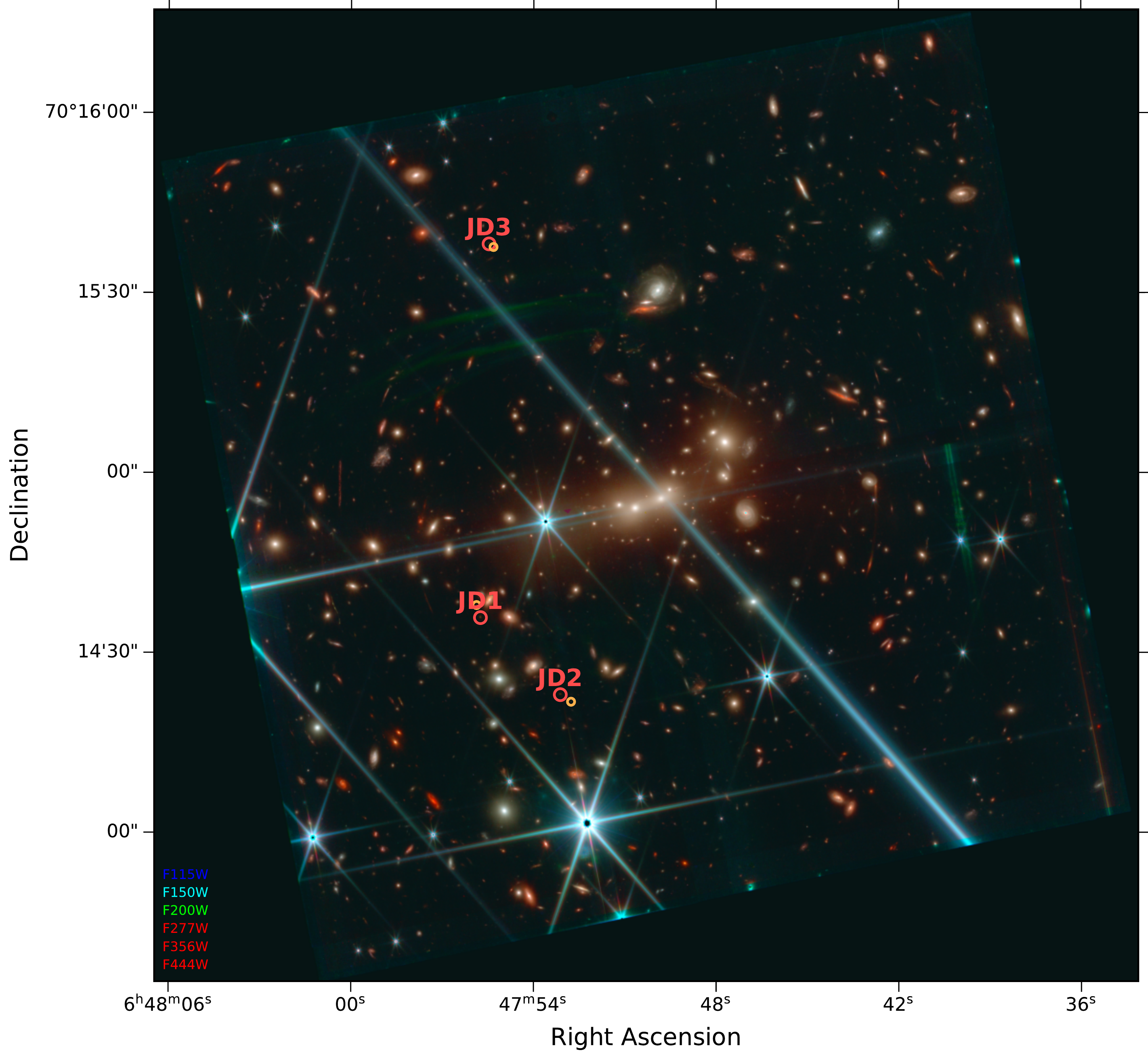}
\caption{JWST NIRCam color image of MACS0647+70 
with the three lensed images of MACS0647--JD labeled in red.
Also circled in orange are the lensed images of a candidate companion galaxy C.
\label{fig:macs0647}
}
\end{figure*}

\section{Observational Data}
\label{sec:data}


We analyze new \JWST\ NIRCam images (GO 1433, PI Coe)
shown in Figure \ref{fig:macs0647},
as well as archival \HST\ images described below
and detailed in Table \ref{tab:obs}.
All of the data are publicly available in the 
Mikulski Archive for Space Telescopes (MAST; DOI: 10.17909/d2er-wq71).
We also provide reduced data products aligned to a common pixel grid (\S\ref{sec:reduction}).


\subsection{HST observations}
\label{sec:HST}
MACS0647+70 has been observed with 39 orbits of \HST\ imaging in 17 filters.
It was first observed by programs
GO 9722 (PI Ebeling) and GO 10493, 10793 (PI Gal-Yam)
in the ACS F555W and F814W filters.
Then CLASH (GO 12101, PI Postman) obtained imaging in 15 additional filters
with WFC3/UVIS, ACS, and WFC3/IR, spanning 0.2--1.7$\mu$m.
Additional imaging in WFC3/IR F140W was obtained as part of a grism spectroscopy program (GO 13317, PI Coe).

\subsection{JWST observations}
\label{sec:JWST}
Here we present new \JWST\ NIRCam imaging 
in 6 filters, F115W, F150W, F200W, F277W, F365W, and F444W spanning 1--5$\,\mu$m.
These public data were obtained 2022 September 23 
as part of Cycle 1 program GO 1433 (PI Coe).
Total exposure times of 2104\,s in each filter achieve 
5$\sigma$ depths of AB mag 28.0 to 29.0
for small sources ($r = 0.1''$ aperture).
Depths were measured by placing circular apertures in blank regions of the image using the \photutils\ \texttt{ImageDepth} routine\footnote{\url{https://photutils.readthedocs.io/en/stable/api/photutils.utils.ImageDepth.html}}.

In each filter, we obtained 4 dithered exposures 
using INTRAMODULEBOX primary dithers 
to cover the 4-5$\arcsec$ gaps between the short wavelength detectors,
while maximizing image area observed at full depth.
Dithering also mitigates bad pixels and image artifacts, 
while improving resolution of the final drizzled images. 
Each exposure uses the SHALLOW4 readout pattern with ten groups and one integration.

Backgrounds were relatively high that time of year for this target
($\sim$80\% higher than minimum).
The telescope was rolled to Position Angle 280$\degree$.
We observed the cluster in NIRCam module A and a nearby ``blank'' field with module B.
The brighter lensed images MACS0647--JD1 and JD2 were observed with NIRCam SW detector A3,
while JD3 was observed with A1.

This program GO 1433 will obtain additional public data 
expected in January 2023:~NIRCam imaging in F200W and F480M 
and NIRSpec MSA PRISM spectroscopic observations.

\input{observations}

\subsection{Data Reduction}
\label{sec:reduction}

We process imaging data from MAST from all the programs above.
The reduced images, along with source catalogs,
are publicly available online\footnote{\url{https://s3.amazonaws.com/grizli-v2/JwstMosaics/v4/index.html}}
along with public data from other \JWST\ programs.

We retrieve the individual calibrated exposures 
processed by the \HST\ and \JWST\ pipelines
(\texttt{FLT} and Level 2b \texttt{CAL} images, respectively).
We then process all of these using the \grizli\ pipeline\ \citep{Grizli},
co-adding all exposures in each filter,
and aligning all stacked images to a common $0.04''$ pixel grid
with coordinates registered to the GAIA DR3 catalogs \citep{Gaia_EDR3}.
The NIRCam short wavelength images are drizzled to $0.02''$ pixels
(on the same pixel grid supersampled $2\times$)
to provide the highest possible resolution.
We leave out the bluest filter \HST\ WFC3/UVIS F225W image,
which contains very few sources making it difficult to process.

We make use of the latest NIRCam calibrations \texttt{jwst\_0995.pmap}, 
based on data from NIRCam CAL program data
and made operational 2022 October 6.
These were not available at the time of processing,
so we recalibrate our data in several steps.
The \JWST\ pipeline used NIRCam calibrations
first made available July 29 in \texttt{jwst\_0942.pmap};
these were the first photometric calibrations based on in-flight data.
Subsequently in late August, updated NIRCam calibrations\footnote{\url{https://zenodo.org/record/7143382}}
were calculated independently
and utilized in \grizli\ v4 image processing of many public datasets, including this one.
These calibrations are consistent to within $< 5\%$ of the most recent
\texttt{jwst\_0995.pmap} calibrations in each filter and detector.
We measure photometry (\S\ref{sec:photometry})
in the \grizli\ v4 images,
then finally apply the necessary flux corrections (Table \ref{tab:recalibration})
to JD1 and JD2 observed in NIRCam detectors A3 and A5,
and JD3 in detectors A1 and A5.


The \grizli\ pipeline applies corrections for $1/f$ noise striping
and masks ``snowballs''\footnote{\url{https://jwst-docs.stsci.edu/data-artifacts-and-features/snowballs-artifact}} caused by high-energy cosmic rays.
It also corrects for stray light features known as ``wisps'' 
that are static and have been modeled in the A3, B3, and B4 detectors in F150W and F200W images.\footnote{\url{https://jwst-docs.stsci.edu/jwst-near-infrared-camera/nircam-features-and-caveats/nircam-claws-and-wisps}}

Finally, the \grizli\ pipeline combines all images in each filter,
drizzling them to a common pixel grid using \astrodrizzle\ \citep{MultiDrizzle,DrizzlePac}.
The NIRCam short wavelength images F115W, F150W, and F200W are drizzled to 0$\farcs$02 pixels,
and all other images are drizzled to 0$\farcs$04 pixels (on the same grid at half the resolution).
All \HST\ and \JWST\ images are aligned to a common world coordinate system (WCS)
registered to the GAIA DR3 catalogs \citep{Gaia_EDR3}.
We create color images  using \texttt{Trilogy}\footnote{\url{https://github.com/dancoe/trilogy}}
\citep{Coe12_Trilogy}.

\input{recalibration}

\subsection{\JWST\ Stellar Diffraction Spikes and Scattered Light Artifacts}
\label{sec:starlight}

At relatively low Galactic latitude $b = 25\degree$,
there are many stars affecting the image.
One particularly bright $\sim$~8th magnitude star $\sim$~2$\arcmin$ southwest of JD1 and JD2
(observed in module B) produces a diffraction spike that crosses the entire module A image of the cluster.
Fortunately, none of the lensed images JD1, 2, 3 are impacted by the spikes,
with the possible exception of one that comes close to JD2 in F277W.

Other scattered light artifacts are isolated and do not impact the lensed images of MACS0647--JD.
``Claws''\footnote{\url{https://jwst-docs.stsci.edu/jwst-near-infrared-camera/nircam-features-and-caveats/nircam-claws-and-wisps}} are visible as horizontal stripes in our F200W image well south of JD3.
These are presumably due to an extremely bright ($K \lesssim 3$ Vega mag) star
very far from the field of view ($10\degree$ in the telescope V3 direction).
They do not move significantly between dithers and cannot be modeled or subtracted.

Dragon's Breath Type II\footnote{\url{https://jwst-docs.stsci.edu/jwst-near-infrared-camera/nircam-features-and-caveats/nircam-dragon-s-breath-type-ii}} is visible as vertical stripes in our F200W image, near the west edge,
extending south of center in the A4 detector, also far from JD1, 2, 3.

\begin{figure*}
\includegraphics[width=\textwidth]{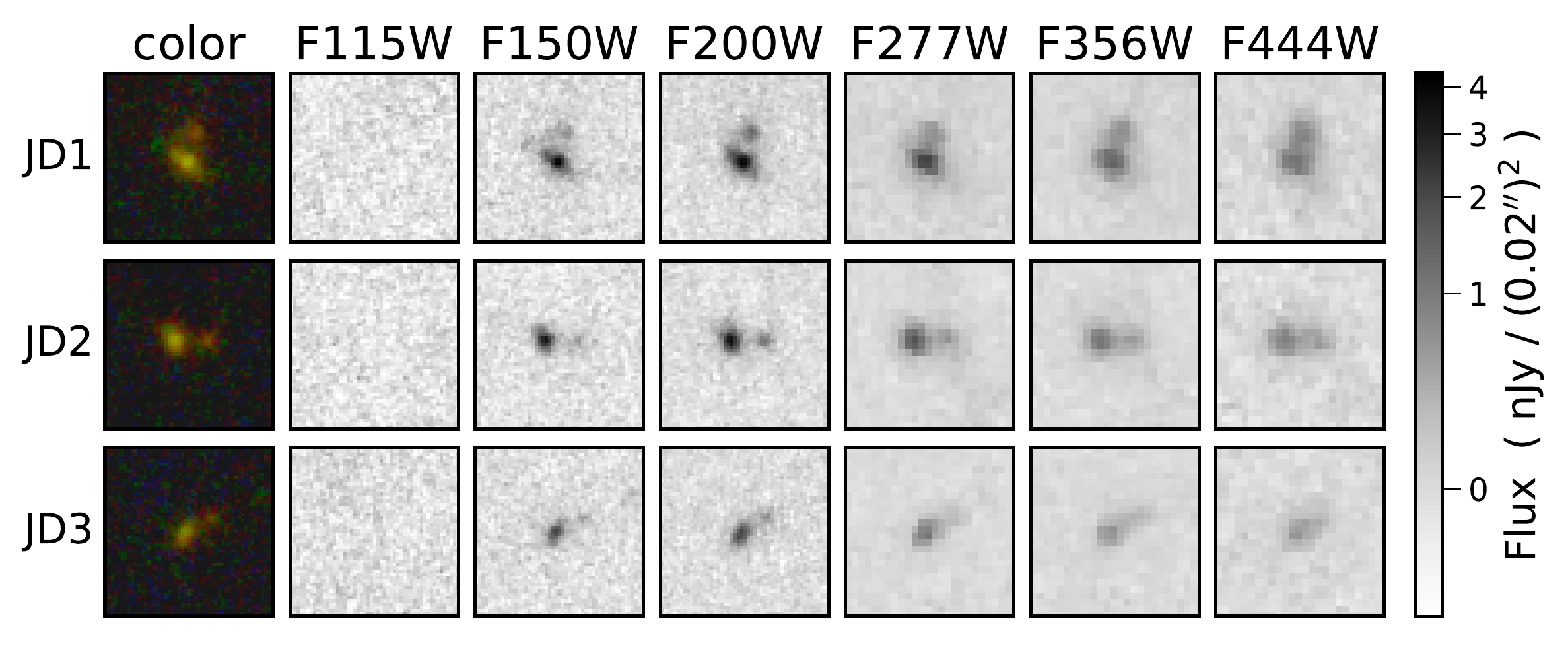}
\caption{MACS0647--JD observed in the NIRCam SW color image
(B=F115W, G=F150W, R=F200W) and in each NIRCam filter.
Each image stamp is $1\arcsec$ across.
Grayscale images for all filters are scaled to the same flux densities 
per $0\farcs02$ pixel, as shown in the colorbar.
Data are recalibrated to \texttt{jwst\_0995.pmap}.
\label{fig:MACS0647--JD}
}
\end{figure*}
\begin{figure*}
\includegraphics[width=\textwidth]{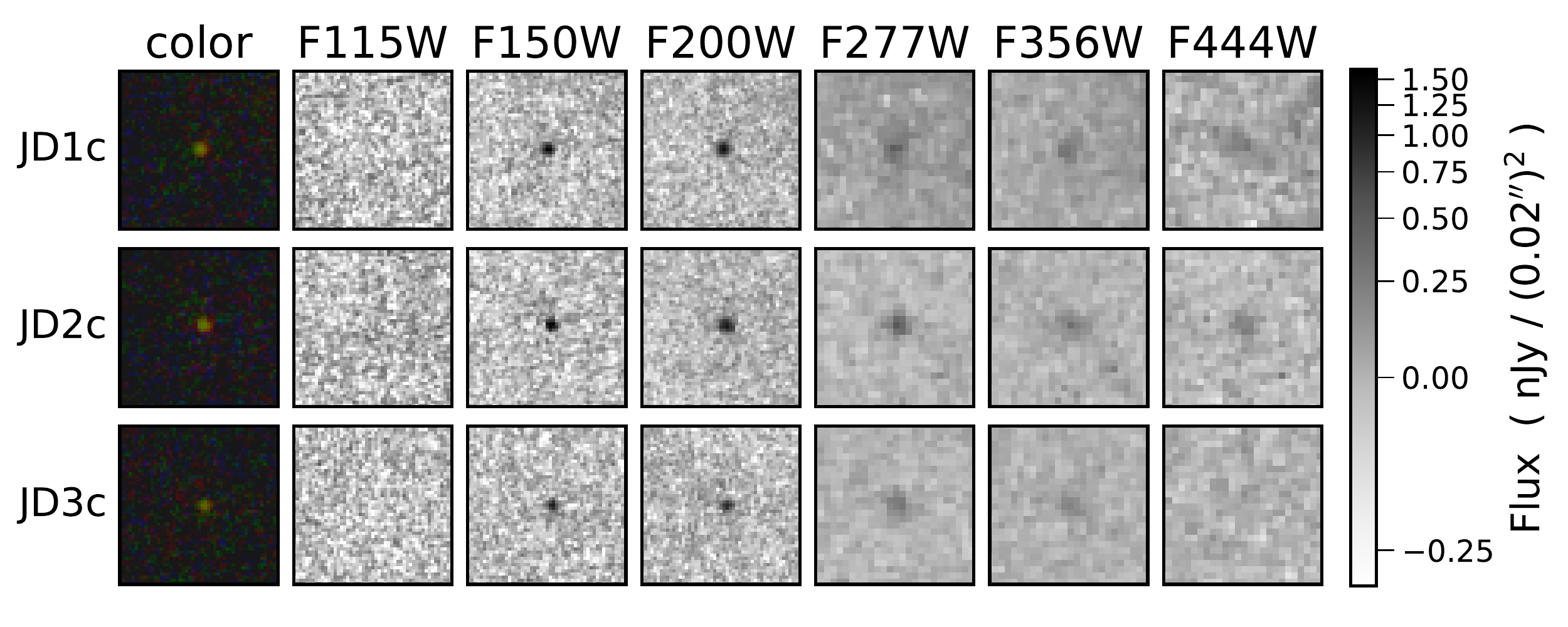}
\caption{Candidate companion galaxy MACS0647--JDc.
Color and grayscale images as in Figure \ref{fig:MACS0647--JD}.
\label{fig:MACS0647--JDc}
}
\end{figure*}

\section{Three Stellar Components}
\label{sec:galaxies}

The \JWST\ NIRCam images clearly resolve MACS0647--JD into two galaxies or components: A and B
(Figure \ref{fig:MACS0647--JD}).
Component A is brighter and spatially extended, 
while B is fainter, more compact, and redder in the short wavelength filters.
These two components are clearly seen in each of the three lensed images JD1, 2, 3.
Both are J-band dropouts, not detected in F115W.

Additionally, we identify a candidate companion galaxy C, another J-band dropout 
(Figure \ref{fig:MACS0647--JDc})
observed 2.2\arcsec, 2.2\arcsec, and 0.9\arcsec\ from JD1A, JD2A, and JD3A, respectively
(see Figure \ref{fig:macs0647}).
It is fainter than A and B and even more compact.

\subsection{Lens Modeling}
\label{sec:lens}

A first lens model for this cluster, prior to CLASH imaging, 
was presented by \cite{Zitrin2011}. 
Lens modeling enabled by the CLASH \HST\ images has been presented in
\cite{Coe2013,Zitrin15_CLASH,Chan17}
using various methods:~Lenstool \citep{JulloLenstool07,JulloLenstool09}, 
Zitrin-LTM \citep{Zitrin09,Broadhurst05}, and
WSLAP+ \citep{Diego05wslap,Diego07wslap2}.
Magnification estimates range from 6.0--8.4, 5.5--7.7, and 2.1--2.8 for JD1, 2, 3, respectively.
Uncertainties are thus roughly $\pm17$\%, 
similar to performances modeling simulated lenses with excellent constraints.
These models have decent constraints with between 9 and 12 multiply-lensed galaxies,
however none have spectroscopic redshift.

\JWST\ imaging reveals more multiply-lensed image systems which will be published alongside a new lens model in Meena et al.~(in preparation).
The model was obtained using a revised, analytic version 
of the parametric method presented in \cite{Zitrin15_CLASH} 
and was recently used, for example, in \citet{Pascale2022}. 
This preliminary, new parametric lens model yields magnification estimates
of $\sim$6.9, 6.3, and 2.1 for JD1, 2, 3,
with tangential (linear) magnifications of $\sim$4.7, 4.4, and 1.8.
Another preliminary new mass model using GLAFIC \citep{Oguri10} predicts magnifications 
of $\sim 9.1$, $5.5$, and $1.8$ for JD1, 2, 3, respectively.

\JWST\ imaging also yields direct new measurements of the observed flux ratios
$\sim$~3.5 : 2.3 : 1 for JD1, 2, 3 based on NIRCam photometry measured in F200W and redward, averaging 340, 223, 97 nJy (AB mag 25.1, 25.5, 26.4).
Based on these measured flux ratios, we adopt fiducial magnification estimates of
$\sim$~8.0, 5.3, and 2.2 for JD1, 2, 3, respectively,
with tangential magnifications of $\sim$~5, 4 and 2.
These are roughly consistent with previous estimates\footnote{{Previous magnification estimates for JD1, JD2, and JD3 were
8, 7, 3 \citep{Coe2013};
6, 6, 2 \citep{Zitrin15_CLASH}; and
8, 8, 3 \citep{Chan17}.
}}
and the total magnification $\sim$~15.5 is roughly equal to the total in our new preliminary lens model.
The average delensed flux in F200W, F277W, F356W, F444W is 43 nJy (AB mag 27.3, $M_{UV}=-20.4$)
with an uncertainty of $\sim$~17\%.

\subsection{Sizes and Separations}
\label{sec:size}

\begin{figure*}
\includegraphics[width=\textwidth]{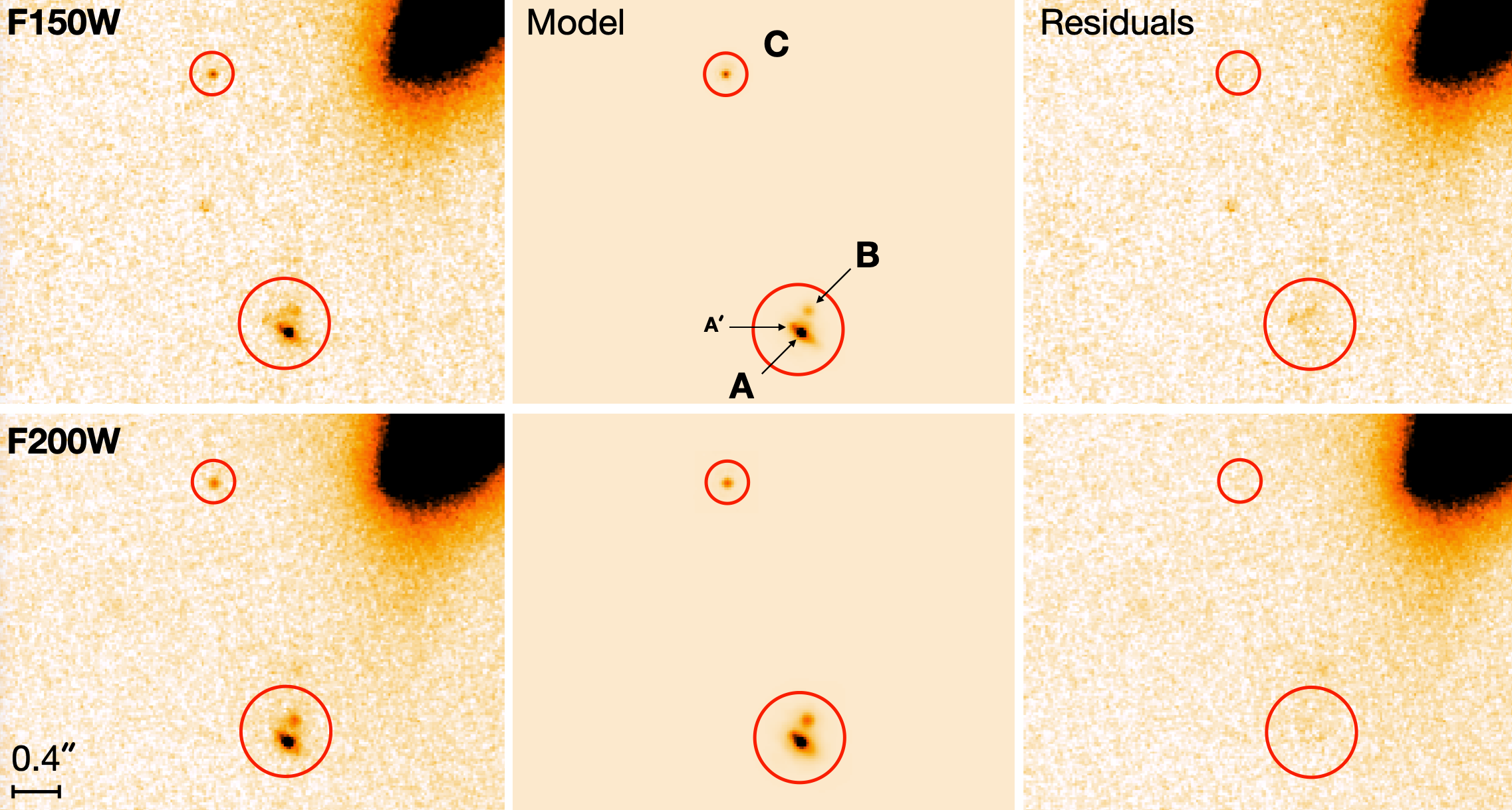}
\caption{
GALFIT modeling (center) of JD1ABC in the F150W and F200W images (left) with residuals shown at right.
S\'ersic model parameters are fit in F150W and held constant (aside from flux) in F200W.
A possible noise fluctuation A$\arcmin$ apparent only in the F150W image
was included as a 4th model component so as not to bias the fitting.
A$\arcmin$ and C are compact and modeled as point sources (convolved with the image PSF).
Based on this model combined with our lens model, 
we measure A and B have radii of $\sim$~70{$^{+24}_{-24}\,$} and 20{$^{+8}_{-5}\,$}pc, respectively (\S\ref{sec:size}).
\label{fig:galfit}
}
\end{figure*}

\begin{figure*}
\includegraphics[width=\textwidth]{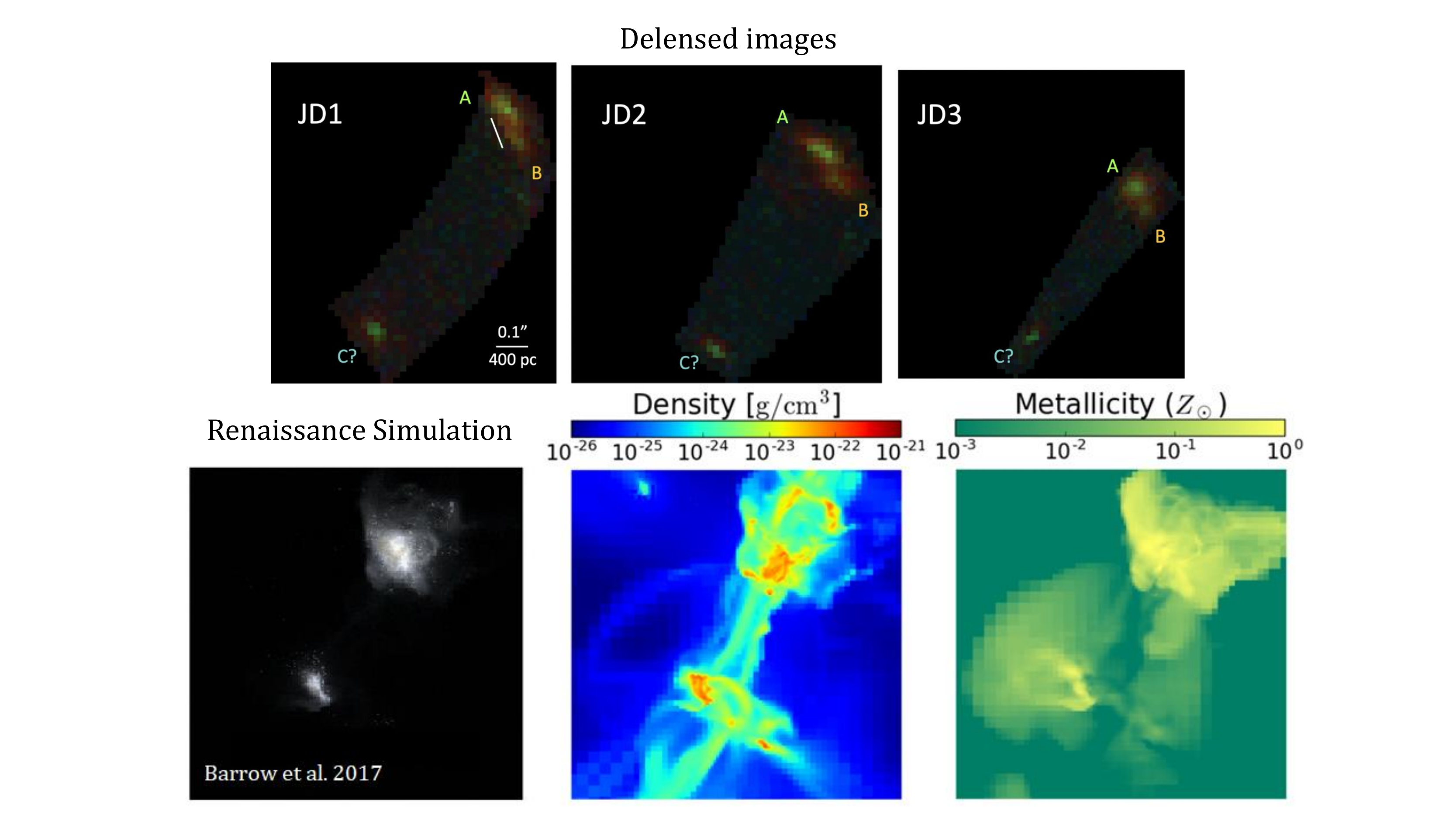}
\caption{Delensed images of JD1, 2, 3 
based on our Zitrin-analytic lens model (Meena et al.~in preparation)
compared to a pair of merging galaxies in the Renaissance simulations \citep{Barrow17}
shown at {bottom on the same physical scale. }
Galaxies A \& B are separated by $\sim 0.1\arcsec \sim 400$ pc in the delensed source plane.
The candidate companion C is $\sim 0.63\arcsec \sim 2.5$ kpc away.
Note the relative positions are reproduced well 
here assuming A, B, and C are all at $z = 10.6$.
The reconstruction is similar for C at $z \sim 9$, with delensed positions changing slightly.
Redshifts $z < 5$ are ruled out for B and C (assuming A is at $z \sim 11$),
as the relative separations would change significantly.
\label{fig:delensed}
}
\end{figure*}

We use GALFIT \citep{Peng2010_GALFIT} to model JD1 A \& B in the sharpest image, F150W.
Galaxy A is fit well by a 2-component S\'ersic model (see Figure \ref{fig:galfit}), including a compact core and a more extended host with a radius of $4.4\pm1.5$ pixels = $0.09\arcsec$ (adopting a Gaussian profile for both, S\'ersic index n=0.5).
De-lensing that by the tangential linear magnification $\sim$~5, yields a radius $\sim$~70{$^{+24}_{-24}\,$}pc.
Galaxy B is well-fitted by a single compact source with a radius of 1.3$^{+0.5}_{-0.3}$ pixels, with a de-lensed radius $\sim$~20{$^{+8}_{-5}\,$}pc.
This analysis method was tested and validated with simulations in \cite{Mestric2022}.
{Note that we use the morphology measurements from F150W to model the F200W image as shown in Figure \ref{fig:galfit} and is well-fitted.}

A similar independent analysis with IMFIT \citep{Erwin2015}
fitting galaxy A to a single component yields a radius of 3 pixels
(with higher S\'ersic index $n \sim 2$ vs.~0.5 for the GALFIT extended component).
This yields a smaller delensed radius $\sim$~45 pc for A.

A third analysis measuring the curve of growth (flux vs.~radius)
yields delensed effective radii $\sim$~70 and 50 pc for A and B, respectively.

To measure separations between A, B, and C, we delens images JD1, 2, 3 to the source plane
(Figure \ref{fig:delensed}).
We find the cores of A and B are separated by $\sim 0.1 \arcsec (\sim 400$ pc)
in both the Zitrin-analytic GLAFIC models.
The candidate companion C is $\sim$~3 kpc away.

\section{Photometry Measurements}
\label{sec:photometry}

\input{photometry}

The \grizli\ pipeline uses SEP \citep{SEP}, a Python implementation of SourceExtractor \citep{SourceExtractor},
to detect sources in a stacked NIRCam image and measure aperture-matched photometry in all filters.
Photometry is measured in circular apertures with radii $0.5\arcsec$.

JD1, 2, 3 are detected as objects \#3593, 3349, 4871 in the public v4 catalog (\S\ref{sec:reduction}).
Table \ref{tab:pho} provides their measured coordinates 
and photometry recalibrated using Table \ref{tab:recalibration}.
These are total fluxes measured for galaxies A+B.
To measure photometry of these components individually, we use the methods below.
The photometry of individual galaxies A and B is organized in Table \ref{tab:phoindi}.

The candidate companion galaxy C is detected as objects \#3621, 3314, 4858.


\subsection{\piXedfit}
\label{piXedfit_phot}

\input{photometry_individual}

To spatially resolve the SEDs of the two galaxies, we use \piXedfit \citep{Abdurrouf2021}. For this resolved SED analysis, we use 13-band imaging from ACS and NIRCam
(excluding ACS F850LP and WFC3/IR filters with broader PSF,
as well as the lower wavelength WFC3/UVIS filters), similar to the analyses carried out in \citet{2023Abdurrouf}.
First, all images are resampled to $0.02''$ pixels using \tt{reproject} \citep{Robitaille2020}. 
Then, we use SEP \citep{SEP} to detect objects in the NIRCam images,
generating a segmentation map defining pixels belonging A+B.

Photometry is measured in elliptical apertures defined within the segments and without overlap.
Aperture A is an ellipse with semi-major axis 0.2$\arcsec$ and semi-minor axis 0.1$\arcsec$.
Aperture B is a circle with radius 0.1$\arcsec$.
Radial profiles decrease within these apertures, reaching a minimum between them that defines their boundary.

We initially forgo PSF-matching to retain spatial resolution.
However, we note measured colors may be affected by lost and/or blended flux
in the redder filters.
The F444W PSF FWHM is 0.14$\arcsec$ with 54\% encircled energy within $r = 0.1\arcsec$,
compared to 70\% for F150W.\footnote{\url{https://jwst-docs.stsci.edu/jwst-near-infrared-camera/nircam-performance/nircam-point-spread-functions}}
We perform aperture corrections based on point source encircled energy
and discuss how this affects results below.
The effect is to make colors redder, though we note this may be an overcorrection 
with flux also blending between A and B.

Aside from A and B, there are no other nearby objects affecting the photometry.
Local backgrounds are small, consistent with zero, and not subtracted.



\subsection{IMFIT}
IMFIT \citep{Erwin2015} has been used to perform 2D fitting to MACS0647-JD. The PSF used in the fitting has been generated, for each filter, using isolated stars. The two clumps have been fitted separately, alternately masking them, followed by a simultaneous fitting step with the parameters for clump A kept fixed. A S\'ersic profile has been used for clump A, while for clump B both S\'ersic and pointlike profiles resulted in similar values for the reduced $\chi^{2}$.
Photometry has then been performed on the models generated from the results from the 2D fitting, using an elliptical aperture for clump A.


\subsection{CHEFs}

CHEFs \citep[from Chebyshev-Fourier functions,][]{Teja2012} are mathematical orthonormal basis specially designed to model the surface luminous distribution of galaxies. First, a segmentation map is created using SourceExtractor \citep{SourceExtractor} to identify the regions that are dominated by each object. Then, objects are sorted by magnitude and fitted with CHEFs, so the light contribution from the brightest objects is removed previous to the modeling of the fainter objects. As CHEF are orthonormal basis, they can fit any shape, thus recovering all the light even in the case of irregular morphologies. The CHEFs model of each object is calculated in a circular region with radius twice the equivalent radius of the area assigned to the object by the segmentation map. However, the flux is measured up to the radius where the profile of the model either converges to zero or submerges into the sky noise.


\section{SED fitting}
\label{sec:sed}

\input{sedfitting}




We perform spectral energy distribution (SED) fitting with various methods
to estimate the photometric redshift and physical parameters of MACS0647--JD.
The various methods adopt different SED templates and assumptions about physical parameters
summarized in Table \ref{tab:sedfitting}.
We also match the observed clump colors to simulated galaxies
with realistic bursty star formation histories (\S\ref{GAINN}).

\subsection{\texttt{EAZY}}
\label{sec:eazy}

Our public dataset includes SED fitting results from EAZY \citep{Brammer2008}
using recently implemented SFHZ templates with redshift-dependent star formation histories.\footnote{\url{https://github.com/gbrammer/eazy-photoz/tree/master/templates/sfhz}}
EAZY fits non-negative linear combinations of these templates to the observed photometry.
The code is fast, analyzing thousands of galaxies in minutes.
It estimates photometric redshifts 
$z = 10.6^{+0.3}_{-0.2}$,
$z = 10.6^{+0.3}_{-0.4}$,
$z = 10.2^{+0.5}_{-0.6}$ (95\% C.L.)
for JD1, 2, 3, respectively 
(A+B components combined, with F200W AB mag 25.0, 25.5, 26.2).

The fainter companion galaxy C 
(F200W AB mag 28.0, 27.3, 27.8 with large uncertainties)
is also a J-dropout that can be well fit to SEDs at $z = 10.6$
given its larger photometric uncertainties, as we show in \S\ref{sec:companion}.
The photometric redshifts are highly uncertain,
with 95\% confidence ranges
0.5--10.2, 2.2--10.3, and 9.8--11.5 for JD1C, JD2C, and JD3C, respectively.

While EAZY also provides quick estimates of physical parameters,
we turn to other methods to more fully explore the parameter space
and estimate values with uncertainties for the individual clumps A and B.


\subsection{\bagpipes}
\label{Bagpipes}

\bagpipes\footnote{\url{https://bagpipes.readthedocs.io}\label{fn1}} \citep{Carnall2018} 
fits redshift along with a multidimensional space of physical parameters
using the MultiNest nested sampling algorithm \citep{Feroz2008, Feroz2009, Feroz2013}.
We run \bagpipes\ with various sets of assumptions.

We use BPASS v2.2.1 SED templates \citep{BPASS},
importantly including binary stars, resulting in brighter rest-UV flux \citep{Eldridge2020,EldridgeStanway22}.
We use the fiducial BPASS IMF {\tt imf135\_300}: \cite{1993MNRAS.262..545K}
slope $\alpha = -2.35$ between 0.5 -- 300 $M_\odot$
and shallower $\alpha = -1.3$ for lower mass stars 0.1 -- 0.5 $M_\odot$.
This is close to the shallower upper mass slope IMF of \cite{Kroupa2002}.
Metallicities range from (0.0005 -- 2) $Z_\odot$.

We reprocess the templates using 
the photoionization code \textsc{Cloudy} c17.03 \citep{Ferland1998, Ferland2013, Ferland2017}
to include nebular continuum and emission lines.
We generate templates for ionization parameter $U$ ranging between log($U$) = $-4$ to $-1$.



We assume an analytic star formation history (SFH) model ``delayed $\tau$'':
SFR$(t) \propto t \ \exp{(-t/\tau)}$.
SFR rises linearly, then slows before declining exponentially,
unless the free parameter $\tau$ is larger than the formation age (as in our fits), 
in which case there is no decline.


For dust attenuation, we use the \cite{Salim18} parameterization
with slope $\delta$ allowed to vary between 0 (Milky Way) and steeper $-0.45$ (Small Magellanic Cloud; SMC),
and 2175\AA\ bump strength $B$ allowed to vary between 0 and 5
(where the Milky Way has $B = 3$ and SMC has $B \sim 0$).
Young stars (age $< 10$ Myr) residing in stellar birth clouds
experience more dust extinction by a factor $\eta$ in the range 1 to 3.

\subsection{\piXedfit}
\label{piXedfit}

As an independent comparison, we also perform SED fitting using \piXedfit \citep{Abdurrouf2021}. For SED modelling, we use Flexible Stellar Population Synthesis \citep[\texttt{FSPS}\footnote{\url{https://github.com/cconroy20/fsps}};][]{Conroy2009, Leja2017}, initial mass function of \citet{Chabrier2003}, Padova isochrones \citep{Girardi2000, Marigo2007, Marigo2008}, MILES stellar spectral library \citep{Sanchez-Blazquez2006, Falcon2011}, and the two-component dust attenuation law by \citet{Charlot2000}. We assume a parametric SFH model in the form of double power-law. FSPS incorporates \textsc{Cloudy} code for modeling the nebular emission. We model the attenuation due to intergalactic medium using \citet{Inoue2014} model. We assume uniform priors for redshift ($2.0-15.0$), age ($0.01-10.0$ Gyr), $Z$ ($\log(Z/Z_{\odot}): [-2.0,0.2]$), and SFH time scale $\tau$ ($0.1-32$ Gyr). The fitting with double power law SFH has two more free parameters that control the slopes of the rising and falling star formation episodes ($\beta$ and $\alpha$). We assume a uniform prior for these parameters with the range of $10^{-2.0}-10^{2.0}$. For the fitting method, we apply the Markov Chain Monte Carlo (MCMC) and set the number of walkers and steps to be 100 and 1000, respectively.

\subsection{\texttt{Prospector}}
\label{Prospector}
To get an independent comparison with non-parametric SFH models, we run SED fitting using \texttt{Prospector} \citep{Leja2017,Johnson2021}, adopting both constant and non-parametric SFHs. Similar to \piXedfit, this code uses FSPS stellar population synthesis models and \textsc{Cloudy} code to account for the nebular emission. We assume a \citet{Chabrier2003} IMF with mass range $0.1 - 300 ~M_\odot$ the IGM attenuation model of \citet{Inoue2014}. We assume a uniform prior for redshift ($z = 6 - 15$), and log-uniform priors for stellar mass ($5 \leq \log(M / M_\odot) \leq 12$), $V$-band optical depth assuming an SMC dust extinction law \citep[$-3 \leq \log(\tau_{V}) \leq 0.7$;][]{Pei1992}, stellar metallicity ($-2.2 \leq \log(Z / Z_\odot) \leq -0.3$, and we assume that the interstellar gas-phase metallicity is equal to the stellar metallicity), and ionization parameter ($-4 \leq \log(U) \leq -1$). For our constant SFH model, we assume a log-uniform prior on formation age from 1$\,$Myr to the age of the universe at the redshift under consideration. Throughout this process, we remove Ly$\alpha$ from the fitting templates.

The non-parametric SFH models implemented in \texttt{Prospector} are piecewise constant functions in time. We adopt eight time bins spanning from the time of observation to a formation redshift, $z_{\rm form} = 15 - 30$ (uniform prior), where the two most recent bins range from $0 - 3$ Myr and $3 - 10$ Myr and the remaining six are spaced evenly in logarithmic lookback time. We adopt the ``continuity'' prior in \texttt{Prospector}, which tends to weight against sharp changes in SFR between adjacent time bins, though we note that the choice of non-parametric prior can have a significant influence on the inferred physical parameters \citep[e.g.,][]{Leja2019,Tacchella2022,Whitler2022a}. 

\subsection{\texttt{BEAGLE}}
\label{BEAGLE}
We also perform SED fitting using \texttt{BEAGLE} tool \citep{Chevallard2016_BEAGLE}. \texttt{BEAGLE} uses templates by \citet{Gutkin2016} which combines the 2016 version of BC03 with the \textsc{Cloudy} code to incorporate nebular emission. We assume a constant SFH model and fit for age with a uniform prior ranging from 1 Myr to the age of the universe at the redshift under consideration. We adopt the same priors on redshift, stellar mass, $\tau_{V}$, $Z$, and $\log(U)$ as for the non-parametric \texttt{Prospector} models. 
We assume that the total interstellar (dust- and gas-phase) metallicity is equal to the stellar metallicity, but note that \texttt{BEAGLE} self-consistently accounts for the depletion of metals onto dust grains, regulated in part by the dust-to-metal mass ratio ($\xi_d$), which we fix to $\xi_d = 0.3$.

\subsection{\texttt{GAINN}}
\label{GAINN} 
Finally, we identified simulated galaxies with colors similar to those observed for MACS0647--JD
to estimate its redshift and star formation history.
We analyzed detailed ENZO \citep{Bryan2014} 
star-forming radiative-hydrodynamic simulations of the early universe
with synthetic photometry generated by \cite{Barrow20}.
The simulated galaxy redshifts and colors were used as a training set for
the Galaxy Assembly and Interaction Neural Network \citep[GAINN;][]{SantosOlmsted2023}.
Additional details are provided in \S\ref{sec:hydrosims} and results are presented below.


\begin{figure}
\includegraphics[width=\columnwidth]{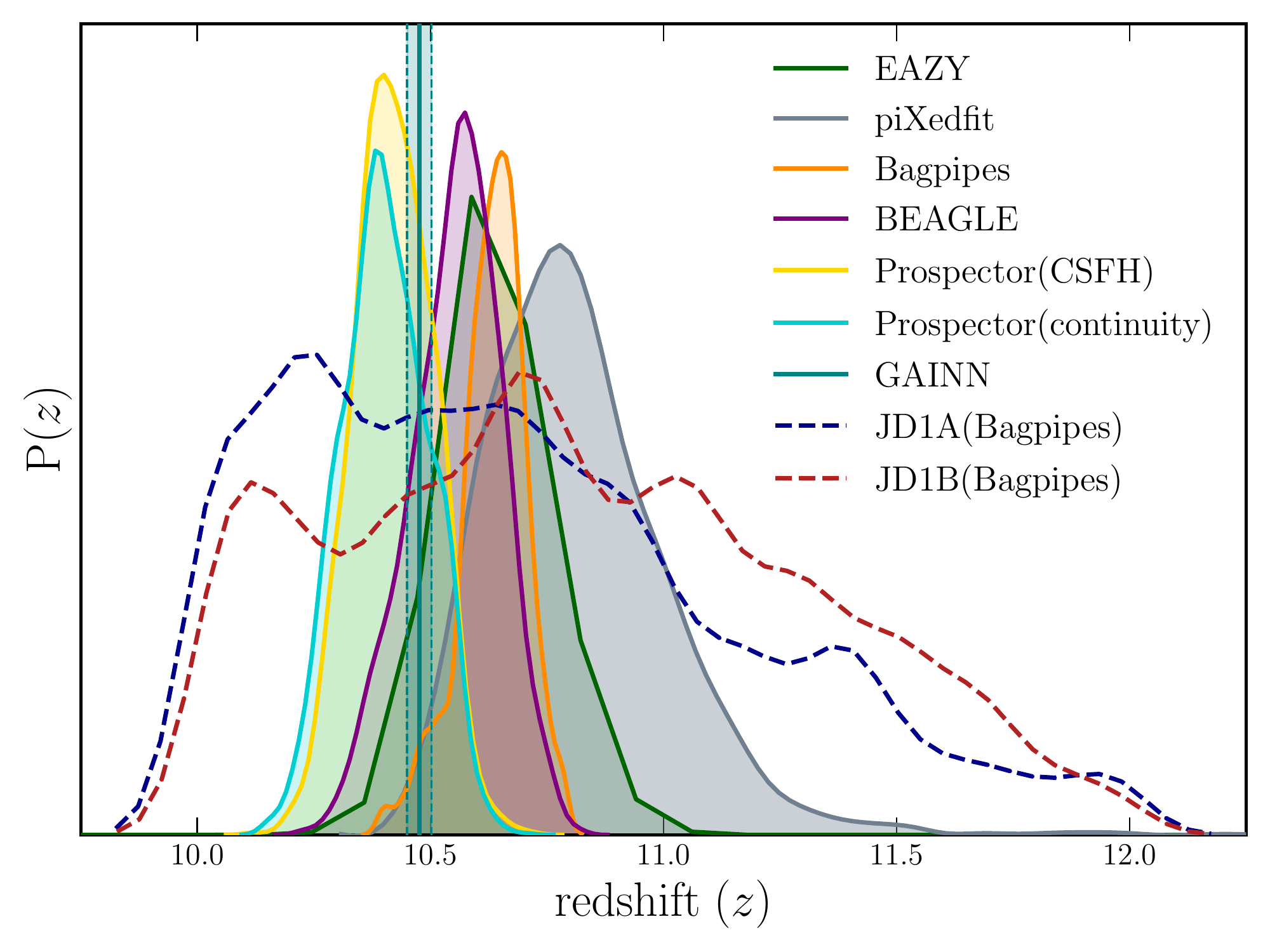}
\caption{Redshift likelihood distributions $P(z)$ of JD1 A+B 
estimated by various methods (\S\ref{sec:photoz}).
Also shown are \bagpipes\ results for the individual clumps A and B.
\label{fig:pz}
}
\end{figure}

\begin{figure*}
\includegraphics[width=\textwidth]{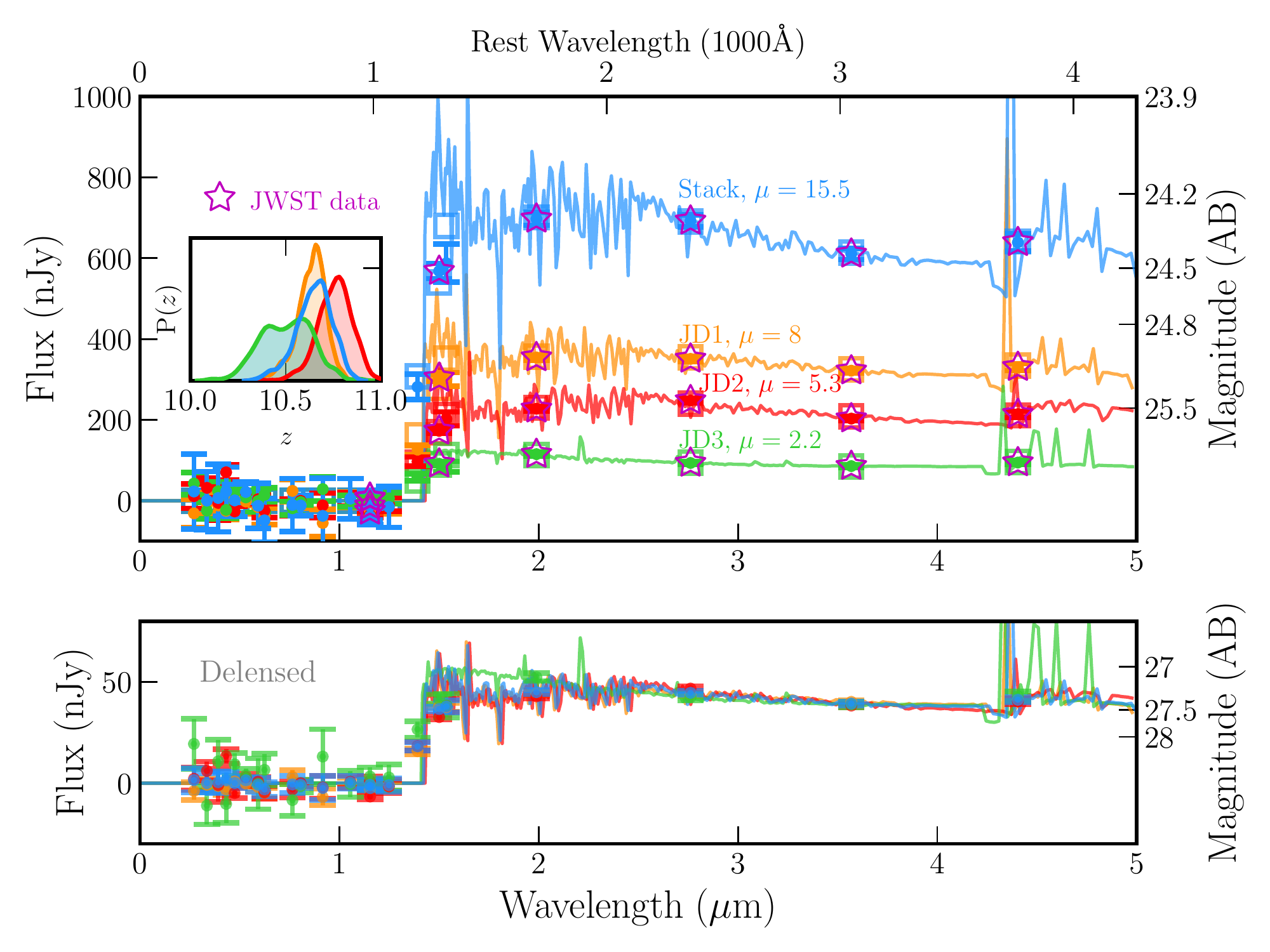}
\caption{
Photometry and \bagpipes\ BPASS SED fits for JD1, 2, 3,
both as observed and delensed by fiducial magnifications $\mu = 8.0$, $5.3$, $2.2$.
Filled circles with errorbars give measured photometry in each filter,
and open boxes show model fluxes for each best fit SED model spectrum shown.
{\JWST} photometry are highlighted by magenta stars.
$P(z)$ of JD1, 2, 3, and the stack one is shown in the small box in the upper panel.
\label{fig:seds}
}
\end{figure*}
\begin{figure}
\includegraphics[width=\columnwidth]{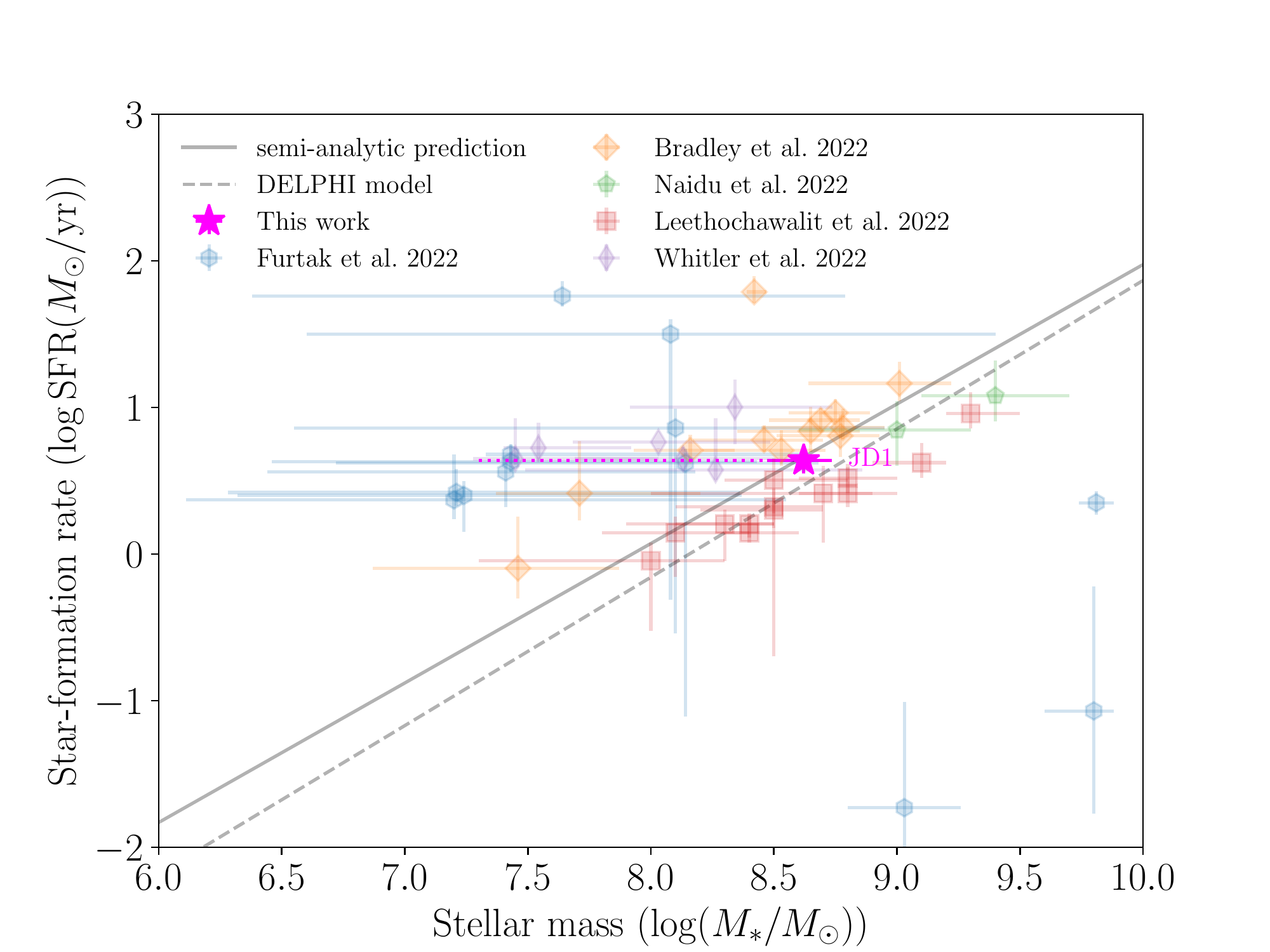}
\caption{
Star formation rate (SFR) vs.~stellar mass for MACS0647--JD
and recently discovered $z \sim 9$ -- 12 candidates analyzed in \JWST\ imaging
\citep{Furtak2022,Bradley2022,Naidu2022,Leethochawalit2022}.
We plot \bagpipes\ results 
for JD1 (A+B).
The results lie along predictions for the $z = 10$ star-formation main sequence
from semi-analytic modeling \citet{Yung2019}
and the DELPHI simulations \citep[]{Dayal2014,Dayal2022}.
{Note that we show the stellar mass estimated from delayed $\tau$ SFH for JD1 (A+B). Possible systematic uncertainty from different assumed SFH is shown in a dotted line.}
%
\label{fig:SFRvsM}
}
\end{figure}
\begin{figure}
\includegraphics[width=\columnwidth]{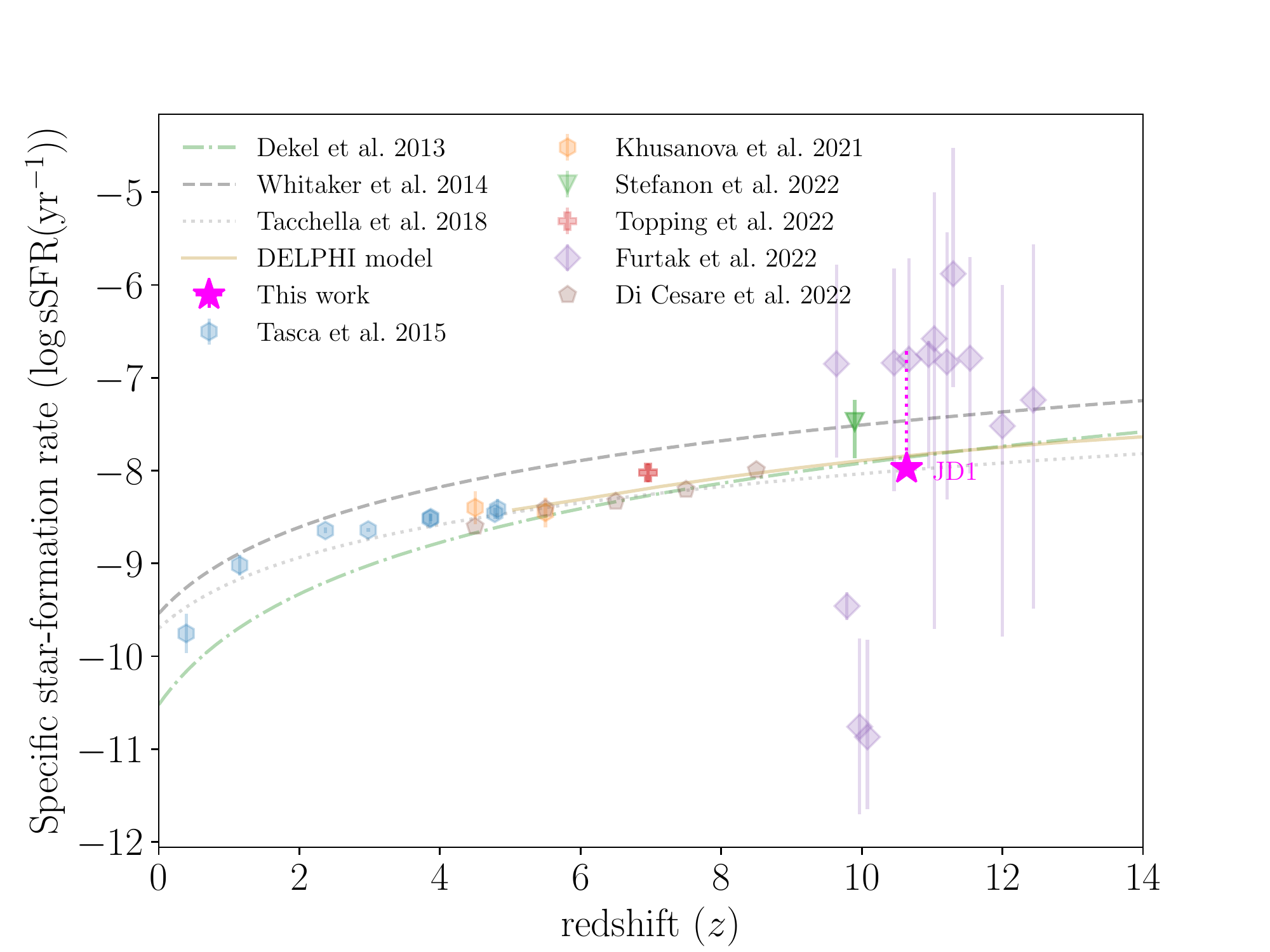}
\caption{Specific star-formation rate as a function of redshift. 
We show the \bagpipes\ result for the JD1 photometry,
while acknowledging the results vary from different methods.
We compare with published results from 
\citet[${\rm log}(M_{*}/M_{\odot})>9.7$]{Tasca2015};
\citet[$9.6<{\rm log}(M_{*}/M_{\odot})<9.8$]{Khusanova2021};
\citet[${\rm log}(M_{*}/M_{\odot})\sim8.4$]{Stefanon2022};
\cite{Topping2022};
\citet{Furtak2022};
and
\citet{DiCesare2022}.
We also compare with predictions from \cite{Dekel2013,Whitaker2014,Tacchella2018}
and DELPHI \citep{Dayal2014,Dayal2022}.
{Note that we show the stellar mass estimated from delayed $\tau$ SFH for JD1 (A+B). Possible systematic uncertainty from different assumed SFH is shown in a dotted line.}%
}
\label{fig:sSFRz}
\end{figure}

\begin{figure*}
\centering{\includegraphics[width=0.8\textwidth]{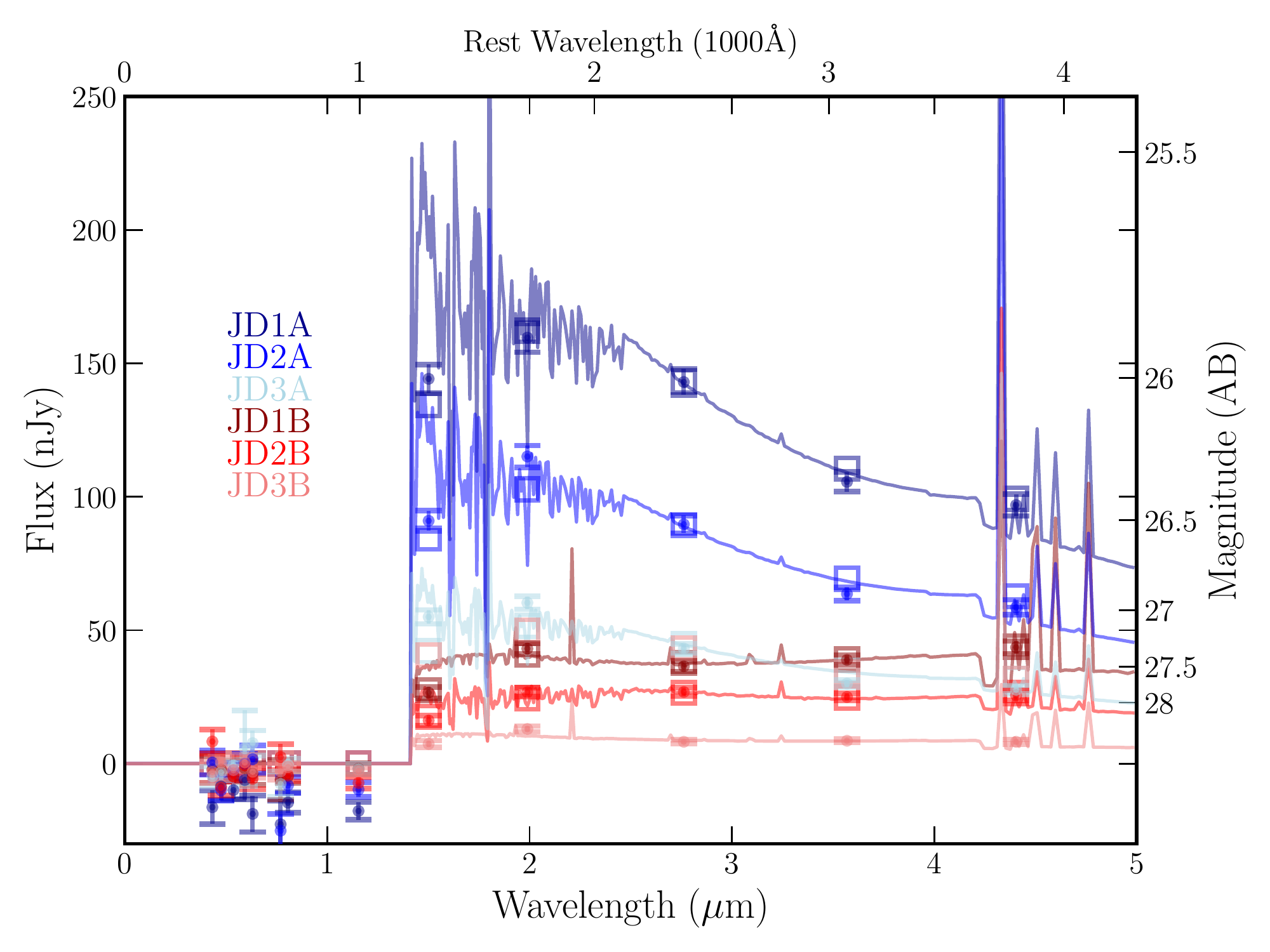}}
\caption{
\bagpipes\ SED fits to photometry of individual clumps A and B
measured by \piXedfit\ with 3\% uncertainty added in quadrature.
\label{fig:sedsAB}
}
\end{figure*}
\begin{figure*}
\centering{
\includegraphics[width=0.49\textwidth]{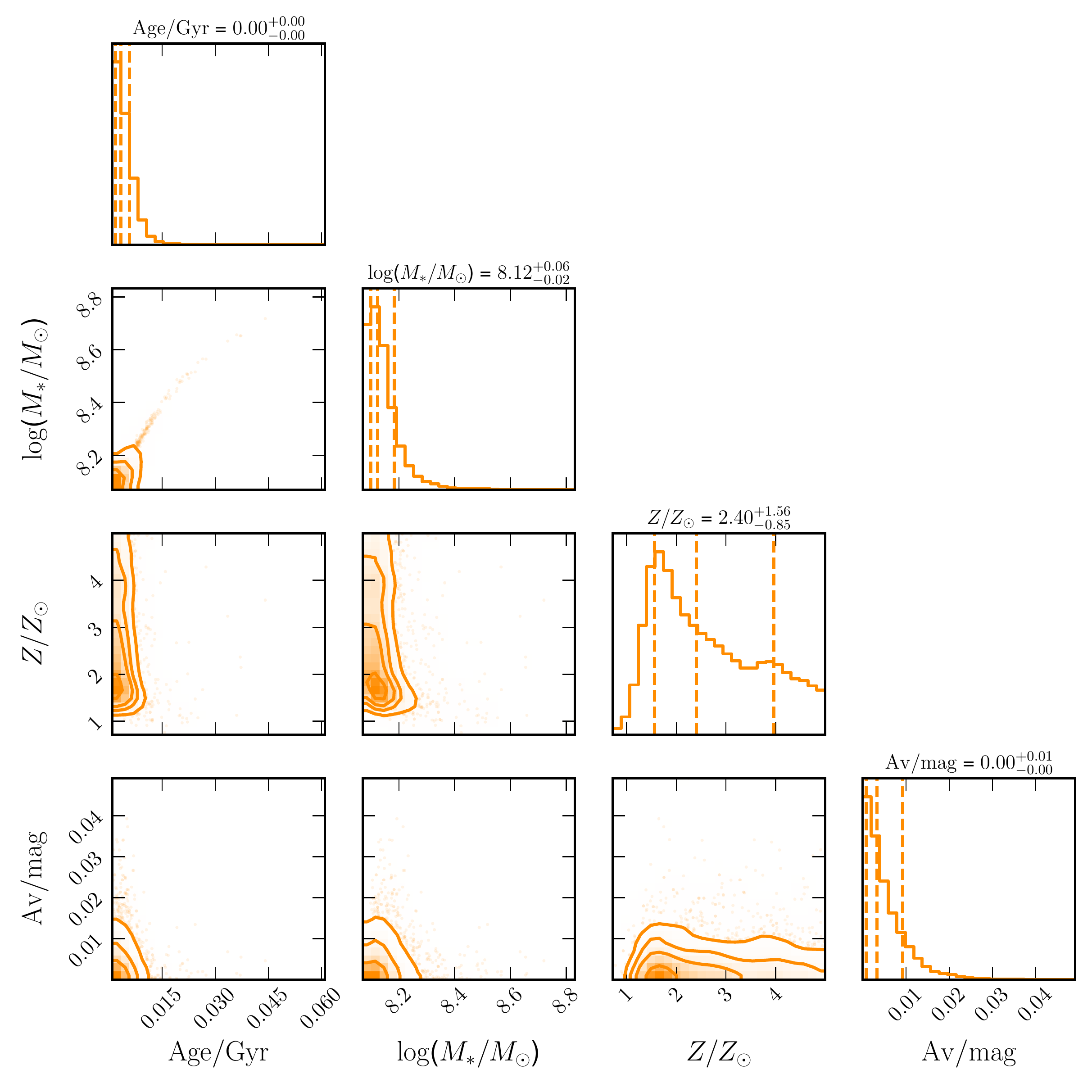}
\includegraphics[width=0.49\textwidth]{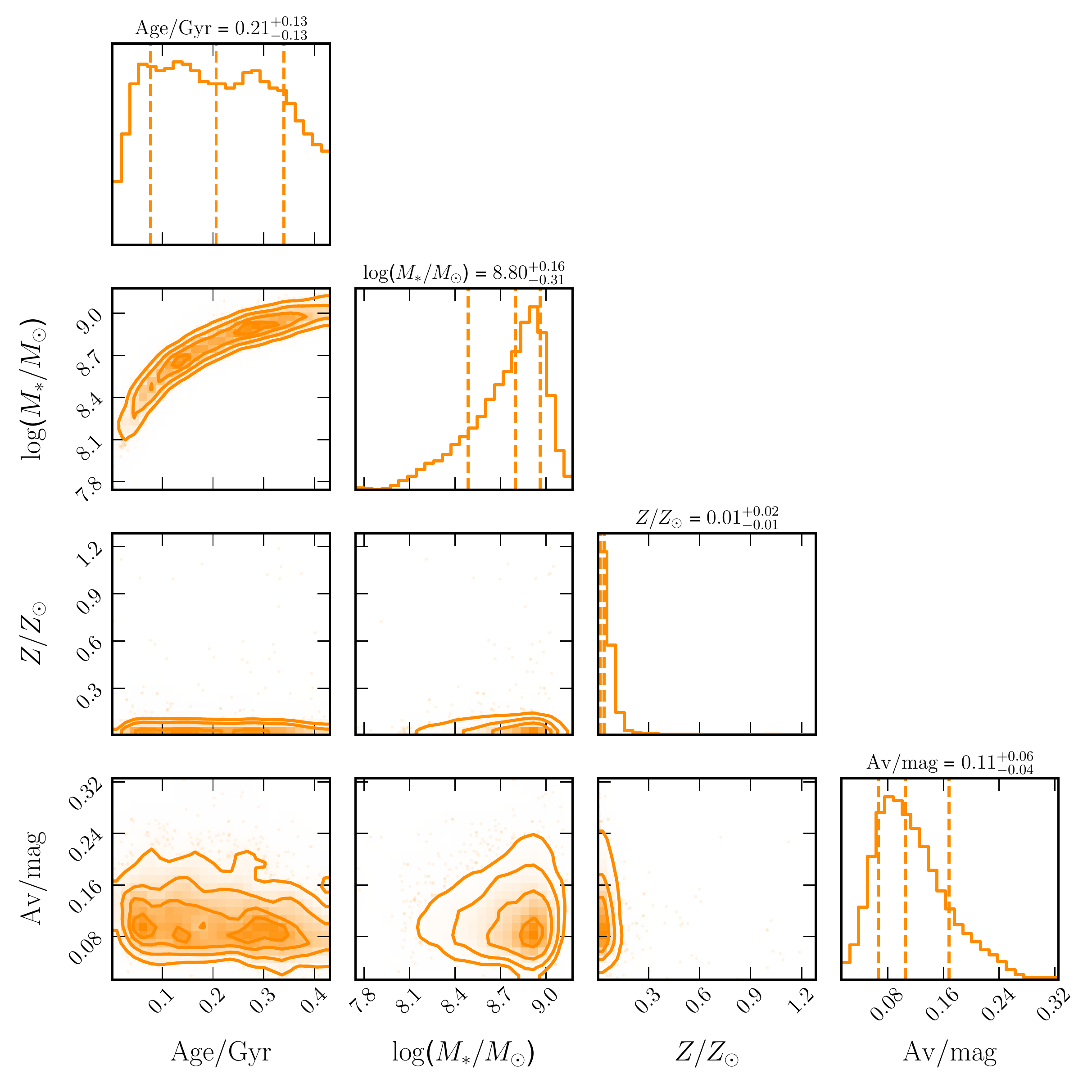}
}
\caption{
{Constraints on age, stellar mass, metallicity, and dust for JD1A (left) and JD1B (right)
from \bagpipes\ with $+3\%$ uncertainty. 
Note that the mass estimates shown here are not delensed.}
\label{fig:corner}
}
\end{figure*}
\begin{figure*}
\includegraphics[width=\textwidth]{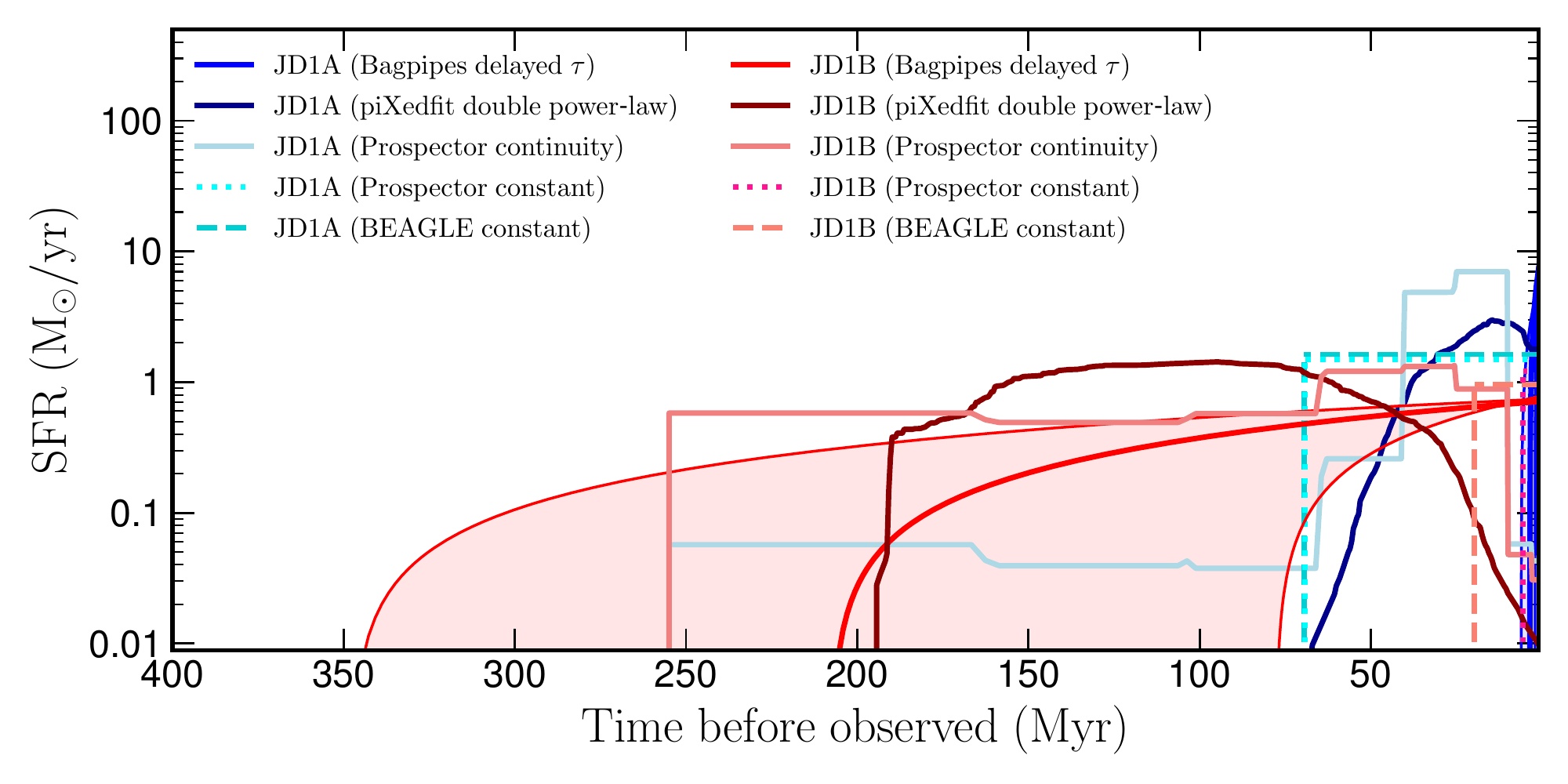}
\caption{
Star formation histories (SFH) of clumps A and B 
for the median ages estimated by each method.
The various SFH parameterizations and assumptions
are summarized in Table \ref{tab:indi_phy}.
For \bagpipes, we plot results from adding 3\% photometric uncertainty.
{Shaded regions indicate the 68 percentile confidence intervals.}
\label{fig:SFH}
}
\end{figure*}
\begin{figure*}
\centering{
\includegraphics[width=\textwidth]{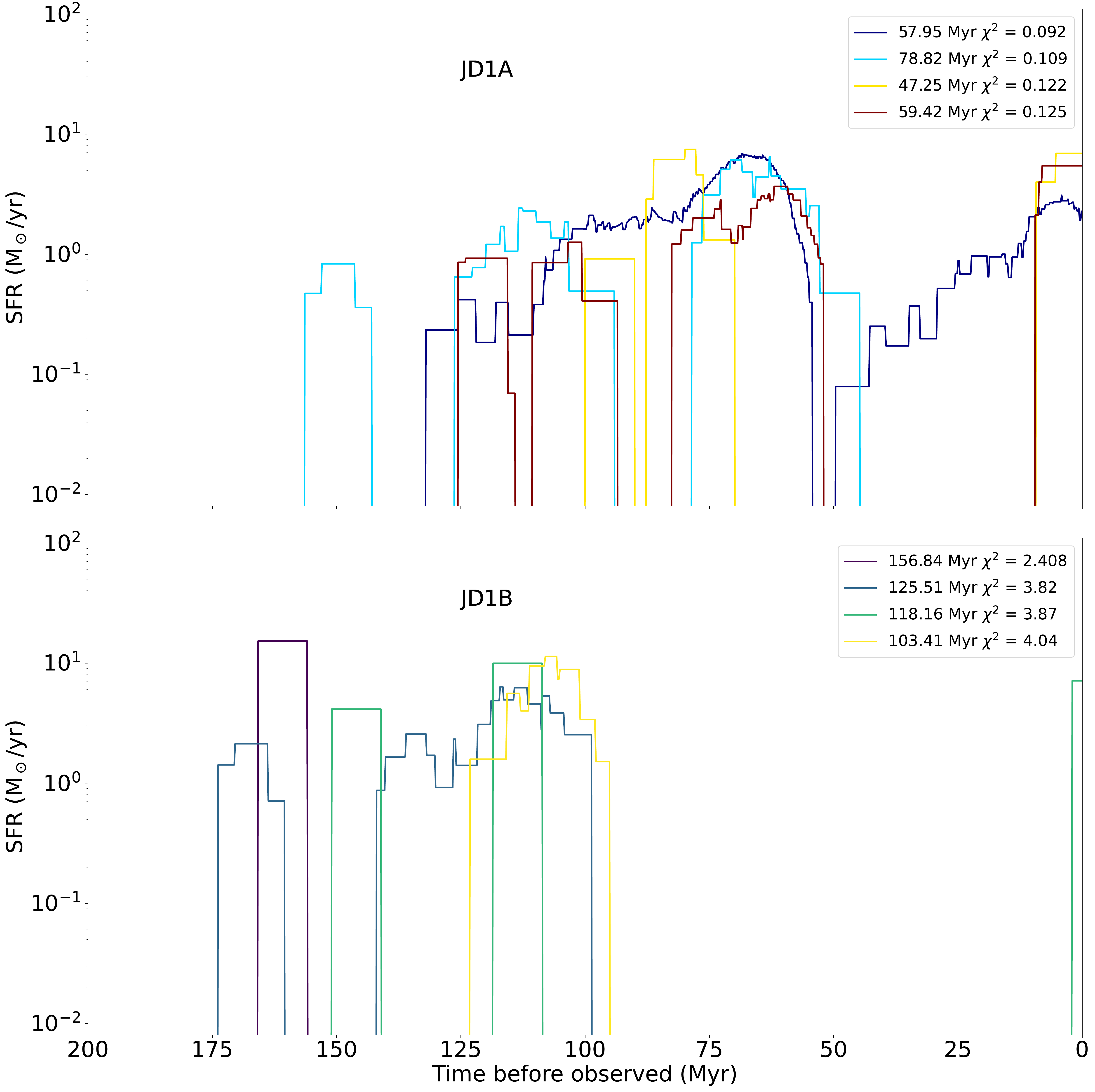}
}
\caption{
Star formation histories of simulated galaxies with colors similar to clumps A (top) and B (bottom).
Note the anti-correlation:
A analogs form most stars while most B analogs are less active,
and vice versa.
Time scales are the same on both $x$ axes.
Mass-weighted ages and $\chi^2$ values from SED fits are given in the legend.
Photometry of the simulated galaxies was measured by \cite{Barrow20}
and matched to the observations using the 
GAINN deep learning network \citep{SantosOlmsted2023}.
\label{fig:simSFH}
}
\end{figure*}

\input{properties}
\input{individual}

\section{Results and Discussion}
\label{sec:results&discussion}

\subsection{Photometric Redshift}
\label{sec:photoz}

MACS0647--JD is confidently at $z = 10.6 \pm 0.3$ (Figure \ref{fig:pz}),
with this range spanning the most likely redshifts from 5 SED fitting packages 
as well as the GAINN deep learning network, 
which estimates $z = 10.48$, 10.81, 10.54 for JD1, 2, 3, respectively.
The components A and B are also each independently strong $z\sim11$ candidates
(with no significant likelihood below $z < 9.5$),
despite lower SNR photometry in each individual object.

We also tried restricting $z < 9$ with \bagpipes,
finding significantly worse ($\chi^2 \sim 500$) fits at $z \sim 0.2$ for JD1, 2, 3
(that reproduce the flat NIRCam colors at 2--5 \um\ but miss the NIRCam F150W and F115W photometry,
also failing to drop out in bluer filters).
Dusty / old galaxies at $z \sim 2$ -- 5 have SEDs that are far too red, 
with SEDs rising through the near-IR.



\subsection{Physical properties} 

In Table \ref{tab:phy}, we report the physical properties of MACS0647--JD treated as a single galaxy
analyzed by \bagpipes\ with photometry from the grizli v4 catalog (recalibrated).
SED fits are shown in Figure \ref{fig:seds}.
We report results for each of the 3 lensed images JD1, 2, 3 and for the stacked photometry,
correcting SFR and mass for magnification.
Assuming A+B had the same star formation history,
this analysis estimates a mass-weighted age $60\pm 25$ Myr,
with SFR $4\pm 1$ $M_\odot$ yr\inv\ averaged over 100 Myr,
stellar mass between 3--6$\times 10^8 ~M_\odot$,
and specific SFR $\sim$~10 Gyr\inv ($\pm10\%$).
We acknowledge the stellar masses estimates are subject to uncertainties
in the star formation history, stellar mass function, 
stellar metallicities, and dust properties.

The SFR and stellar mass are consistent with the predicted stellar main sequence
from semi-analytic models \citet{Dayal2014,Yung2019, Dayal2022}
and simulations \citep{Dekel2013,Whitaker2014,Tacchella2018,Behroozi2019}.
We plot these relations and results from other $z\sim9$ -- $12$ candidates
measured in \JWST\ observations in Figure \ref{fig:SFRvsM}.

The specific SFR is also consistent with predictions at this redshift 
as shown in Figure \ref{fig:sSFRz}.
Note in these model predictions \citep{Dekel2013,Whitaker2014,Tacchella2018,Behroozi2019,Dayal2022}, 
sSFR is relatively flat at high redshifts,
increasing only 0.2 dex from $z = 6$ to 11.
This suggests a significant role for mergers in the early universe;
sSFR(z) would continue to rise more as $\propto (1+z)^{2.25}$
if growth were dominated by cold-mode accretion \citep[e.g.,][]{Dekel2009}.

\subsection{Components A and B: Ages, Dust, and Mass}

In Table \ref{tab:indi_phy}, we report results for components A and B analyzed individually
by various SED fitting methods with photometry from \piXedfit.
The fiducial values are organized in Table \ref{tab:clumps}.
SED fits from \bagpipes\ are plotted in Figure \ref{fig:sedsAB}.
The corner plots of A and B are provided in Figure \ref{fig:corner}.
Star formation histories from SED fitting methods 
are plotted in Figure \ref{fig:SFH}.
Star formation histories from simulated galaxies matching the colors of A and B
using GAINN are plotted in Figure \ref{fig:simSFH}.

B's redder color may be explained by age and/or dust.
Results vary depending on the method and assumptions, including star formation history.
Dust is negligible ($A_V < 0.02$ mag) for A in most analyses,
and slightly higher ($A_V \sim 0.1$ mag) for B,
assuming steep SMC-like attenuation strongly suppressing the rest-UV.


Stellar mass estimates are on the order of $10^8 M_\odot$
with some agreement on higher mass for clump A by a factor of 2 or more
(e.g., A $\sim$$2\times 10^8 ~M_\odot$; B $\sim$$10^8 ~M_\odot$).

Mass-weighted ages from the SED fitting methods
range up to $\sim$~50 Myr and $\sim$~100 Myr for A and B, respectively.
GAINN analog simulated galaxies, similarly,
have mass-weighted ages $50\pm5$ Myr and $125^{+24}_{-12}$ Myr
for A and B, respectively.

B's SED was relatively rare among the simulated galaxies.
It was best matched by galaxies that formed most of their stars over 80 Myr prior to observation,
then either remained less active ($< 0.01 ~M_\odot$ yr\inv) 
or perhaps had some shorter burst of star formation.
The simulated galaxies with colors similar to A had star formation {dissimilar}:
bursty during that period when B was less active.


JD1A is intrinsically very blue {($\beta \sim -2.6\pm0.1$)}
as measured with a power-law fit to the F200W, F277W, and F356W photometry measured by \piXedfit,
where $\beta$ is the rest-frame UV continuum slope
$F_\lambda \propto \lambda^\beta$ (or $F_\nu \propto \lambda^{\beta+2}$).
We measure $\beta = -2.69$ without PSF correction 
and $\beta = -2.54$ after correcting for 
point source encircled energy within $r = 0.2\arcsec$ (see \S\ref{piXedfit_phot}).

Other recent \JWST\ observations have revealed 
even bluer slopes ($\beta \sim -3$) in galaxies at $z \sim 7$ -- 8.5 \citep{Topping2022}
and in a candidate $z \sim 16$ galaxy \citep{Furtak2022},
all with stellar masses of the order of $10^8 ~M_\odot$.
\cite{Topping2022} found these blue colors required large escape fractions
$f_{\rm esc,\HII} \sim 0.6$ -- 0.8
of photons leaking directly from stellar {\HII} regions, bypassing nebular reprocessing.
Our measured {$\beta \sim -2.6\pm0.1$} is slightly redder
and can be fit by our SED models that all assume $f_{\rm esc} = 0$.
Nevertheless, it is in the regime where some significant $f_{\rm esc}$
should be considered to avoid biasing age and mass measurements.

Binary stars are important to include in SED modeling as in BPASS,
especially for such blue galaxies \citep{Eldridge2020,EldridgeStanway22}.
Binary interactions produce more Wolf–Rayet / helium stars at later ages,
generating more energetic photons.
Thus blue observed SEDs may be fit well 
by older (tens of Myr) BPASS templates including binaries, 
whereas templates without binaries may require very young ages ($< 10$ Myr).

We also note uncertainty in the photometry and some variation in $\beta$
measured by the various methods and in the three lensed images.
Ultimately, upcoming NIRSpec spectroscopy will improve measures of $\beta$,
age, and reveal other signatures of large escape fractions.

\input{clumps}

\subsection{Stellar Mass and SFR Densities}

The stellar complexes in MACS0647--JD are very dense with $\sim10^8\,M_\odot$ stellar mass packed into effective radii of $\sim$~70{$^{+24}_{-24}\,$} and 20{$^{+8}_{-5}\,$}\,pc for A and B, respectively.
Assuming fiducial masses 
$10^8$ and $6 \times 10^7\,M_\odot$ 
for A and B, respectively,
the stellar mass surface densities $\Sigma_{\rm eff}$ are roughly on the order of
$\sim$2000 and $\sim$12000\,$M_\odot$ pc$^{-2}$,
where $\Sigma_{\rm eff} = (M / 2) / (\pi R_{M/2}^2)$ 
and the half-mass radius $R_{M/2} = (4/3) R_{\rm eff}$.\footnote{The factor of $4/3$ 
might not be warranted since these are much larger than star clusters,
in which case the densities would increase by a factor of $\sim$~2. See \cite{Portegies-Zwart10}.}
These are higher than the highest density $\sim$1800 $M_\odot$ pc$^{-2}$ reported by \cite{Chen22}
because their size measurements, unaided by lensing,
were only sensitive to structures with radii $> 150$ pc.
The density for clump B is just an order of magnitude less than 
the maximal density $10^5 ~M_\odot$ pc$^{-2}$ reported by \cite{Hopkins10}.

A similar rough estimate of SFR densities yields $\sim$~20 and $\sim$~120\,$M_\odot\,$yr$^{-1}$\,kpc$^{-2}$.
These are high values, 
though less than the highest values $> 1000~ M_\odot$ yr$^{-1}$ kpc$^{-2}$ reported for SMGs.


\subsection{Galaxy clumps or merger?}
\label{sec:merger}

The stellar components A and B may be two merging galaxies,
or they may be two clumps that formed together in situ $\sim400\,{\rm pc}$ apart within a single galaxy.
We cannot distinguish these scenarios with the current data.

Given their masses $\sim 10^8\, M_\odot$, they would affect their surroundings 
such that we would expect them to form at the same time within a galaxy.
A significant age difference would suggest they are separate galaxies now merging.
We cannot conclusively distinguish the ages of A and B given the current data,
though the simulated analog galaxies {at similar redshfit and SED fitting} do strongly suggest star formation {to be dissimilar}
(see \ref{sec:hydrosims}).

MACS0647--JD was discovered in CLASH imaging with a search volume of a few times 1000 Mpc$^3$ =
(10 cMpc)$^3$ at $z \sim 11$ \citep{Coe2013}.
We employ the Astraeus cosmological simulations \citep{Hutter2021}
to calculate the likelihood of finding such a merger within that search volume.
With a box-size of (230 cMpc)$^3$ and a minimum resolved halo mass of $M_h \sim 10^{8.6}~M_\odot$, 
these simulations are ideally suited for such statistics. 
We find 0.176 mergers in a (10 cMpc)$^3$ volume for systems such as JD1 and JD2 
and 0.056 mergers per (10 cMpc)$^3$ 
for three clumps with $M_*$ $\sim$ $10^{8-9}\,M_\odot$ \citep{Legrand2022}.

Dust may also contribute to the different colors observed between A and B.
Clumps are often obscured by different amounts of dust within a galaxy.
Atacama Large Millimeter/submillimeter Array (ALMA) observations of $z \sim 7$ galaxies reveal spatially varying dust
that is sometimes even offset from the stars observed in the rest-UV
\citep{Bowler22,Dayal2022}.
\JWST\ observations of $z \sim 6 - 8$ galaxies in blank fields
reveal clumpy morphologies are common;
each galaxy has a few multi-colored star-forming complexes 
separated by $\sim300$ -- $4300\,{\rm pc}$
with various ages, dust reddening, and masses 
\citep{Chen22}.

The ground-based spectroscopic survey VIMOS UltraDeep Survey (VUDS)
found 21--25\% of $z \sim 2 - 6$ galaxies
are dominated by two massive clumps, each $\sim$$10^9 - 10^{10} ~M_\odot$,
with smaller fractions of galaxies having 3, 4, or more clumps \citep{Ribeiro17}.
Major mergers are invoked to explain the galaxies with two massive clumps,
while disk instability can explain the formation of 3 or more smaller clumps in situ.
 


\begin{figure*}
\includegraphics[width=0.33\textwidth]{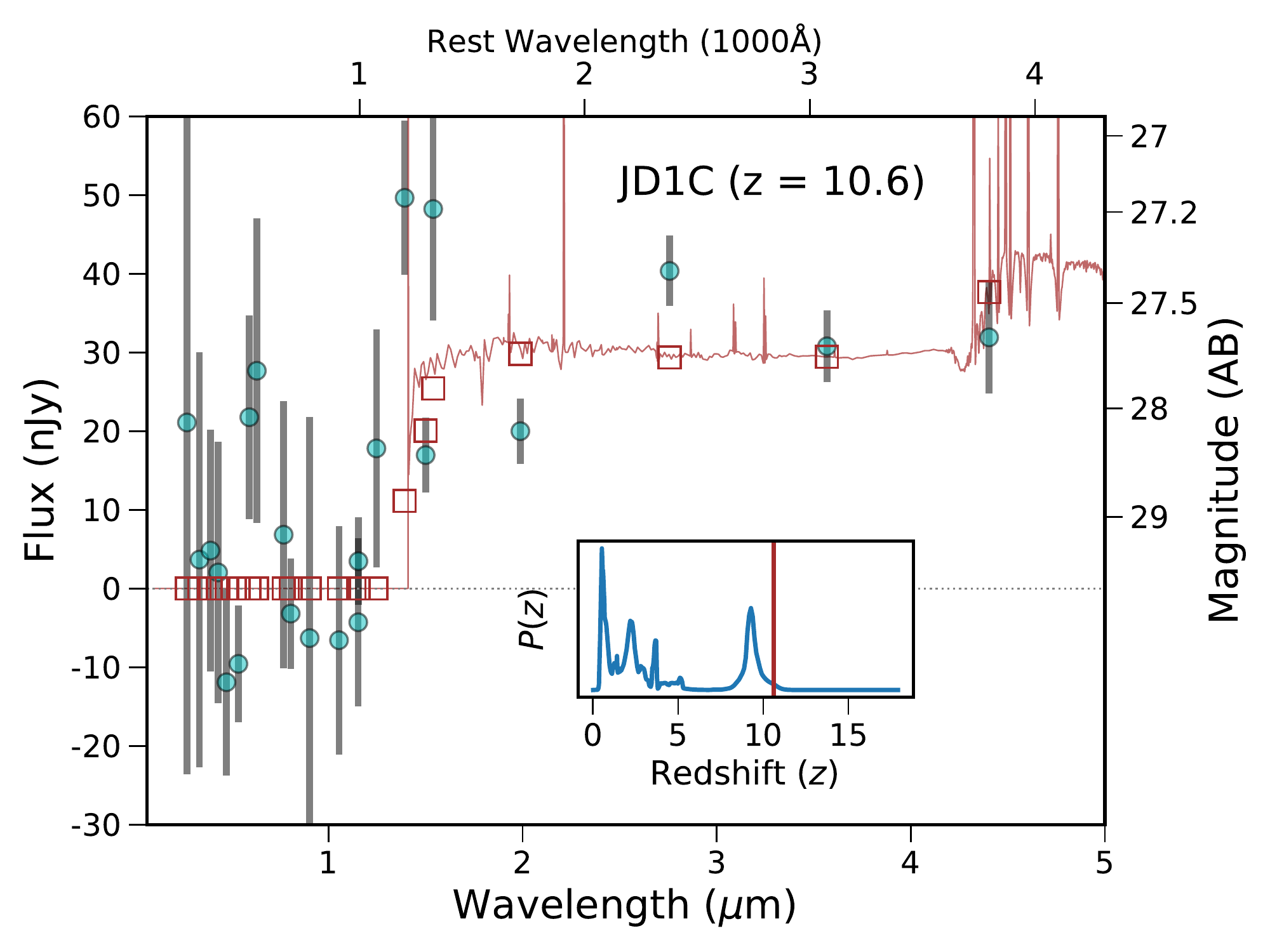}
\includegraphics[width=0.33\textwidth]{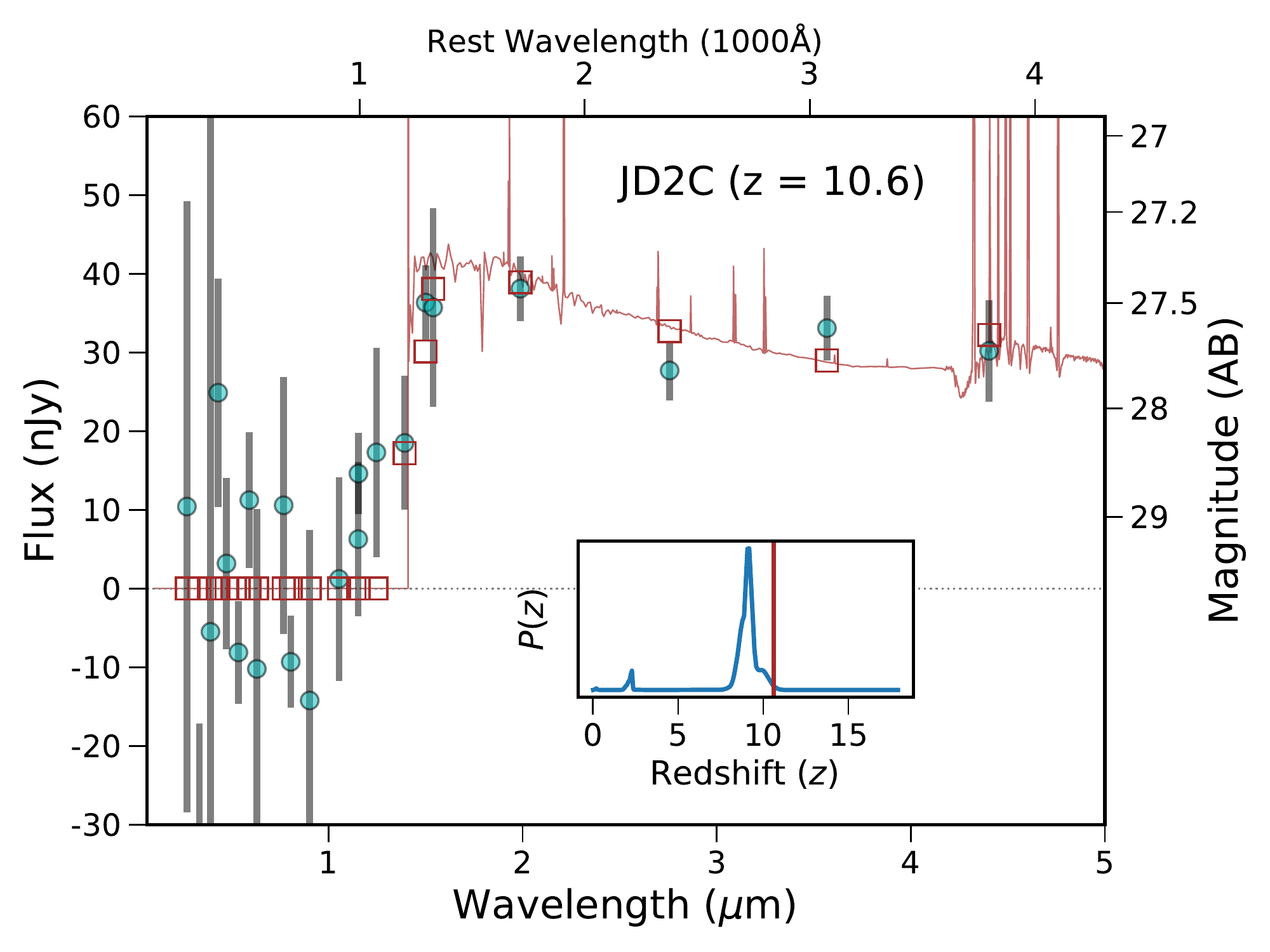}
\includegraphics[width=0.33\textwidth]{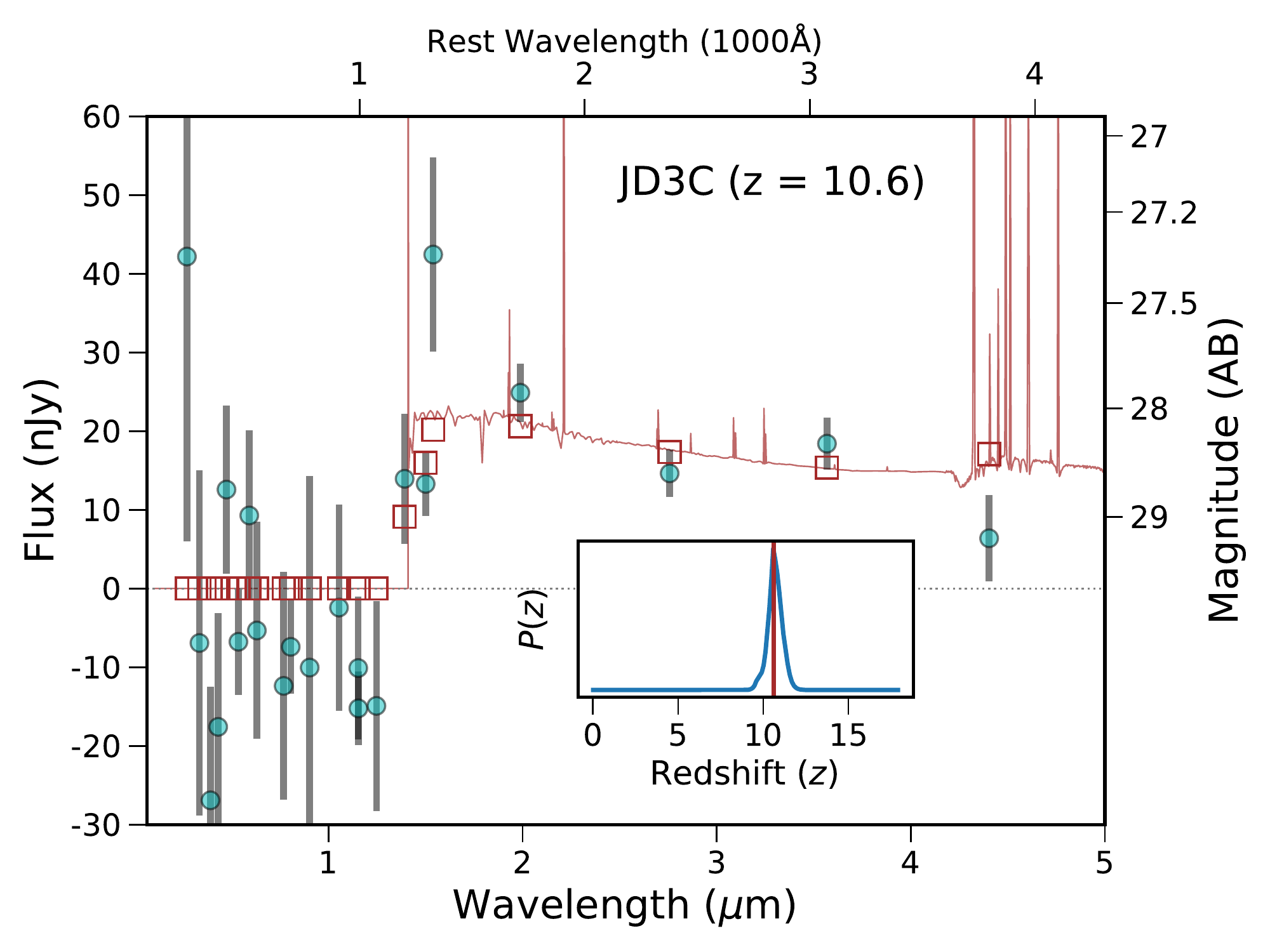}
\caption{Candidate companion C is fainter and the photometric redshift is less clear,
but its photometry can be well fit by SEDs at $z = 10.6$. 
EAZY SED fits are shown for JD1C, JD2C, and JD3C, left to right.
Their redshift likelihood distributions $P(z)$ are shown in the insets
before fixing their redshifts to $z = 10.6$ for the SED fits shown.
\label{fig:EAZYC}
}
\end{figure*}

\subsection{A possible companion}  
\label{sec:companion}

The candidate companion galaxy C is also a J-band dropout
with three lensed images at the predicted locations for a galaxy 3 kpc away at $z \sim 11$.
It is $>2$ mags fainter, so the photometric redshifts are more uncertain.
In Figure \ref{fig:EAZYC}, we show the photometry for JD1C, JD2C, and JD3C
can all be well fit with EAZY SEDs assuming $z = 10.6$.
(This is also the most likely redshift for JD3C, though it is not for JD1C and JD2C.)

The redshift can also be constrained from strong lens mass modeling. 
Both of our lens models (\S\ref{sec:lens}) find the observed lensed image locations
are best fit by high-redshift solutions $z \sim 11$.

\subsection{Comparisons to Simulated Galaxies}


We compare our results to various expectations from large cosmological semi-analytic models (SAMs), an N-body cosmological simulation, and high-resolution zoom-in hydrodynamic simulations of the early universe.

We first compare to the DELPHI semi-analytic model \citep{Dayal2014,Dayal2022}. 
In brief, this model reconstructs galaxy halo assembly histories from $z \sim 40$ -- 4.5,
tracking buildup of gas and star formation, including feedback.
With minimal free parameters and calibration, it reproduces observed high-$z$ luminosity functions, stellar mass functions, and ALMA dust estimates.
Details are provided in \S\ref{sec:DELPHI}.

Given a stellar mass of $10^8$ ($10^9$) $M_\odot$ at $z \sim 10.6$, 
DELPHI predicts 
a host halo mass $M_h \sim 10^{10}$ ($10^{11}$) $M_\odot$, 
absolute UV magnitude $M_{\rm UV} \sim -19.5$ ($-20$), 
dust $A_V \sim 0.03$ (0.12) mag,
and a stellar radius \about~70 (350) pc for galaxies at $z \sim 10$,
consistent with our observations, 
especially for the lower end stellar mass $10^8 ~M_\odot$. 
Finally, this model predicts stellar mass-weighted ages that range between 35--180 Myr for galaxies of a similar mass at $z \sim 10$; the range of ages 
reflects the varied assembly histories.

These small amounts of dust are also consistent with recent modeling by \cite{ferrara2022}
suggesting that negligible dust at these redshifts
could help explain the unexpectedly large numbers of $z \sim 10$ -- 14 candidates reported in early \JWST\ observations.

Next, we consider the merging galaxies from a hydrodynamic simulation \citep{Barrow17}
presented in Figure \ref{fig:delensed} that bears a resemblance to MACS0647--JD A+B and C.
This was the most massive halo in that simulation, and it has a total stellar mass $4 \times 10^7 ~M_\odot$. The analog C is connected by a faint filament of stars and gas. This configuration was likely the result of previous galaxy mergers, including A+B, as evidenced by hot regions tracing supernova remnants.


Within a separate hydrodynamic simulation of a 66 Mpc$^3$ co-moving volume with adaptive mesh refinement resolution sufficient to track gas down to 0.25 pc at $z=12$ and form individual Population III stars as well as metal-enriched star clusters \citep{SantosOlmsted2023}, we perform a more thorough search
for simulated galaxies with colors similar to A and B.
The best-matching analogs have mass-weighted ages of \about~50 and \about~125 Myr,
with B having little or no SFR
within the past 100 Myr before observation.
Bursty SFHs with dormant periods of several 10 Myr are common in these simulations.
Observationally, the question is whether we observe them when they are active with SFR in the past 10 Myr
and thus brightest in the rest-UV.


JD1A's photometry was well matched to SEDs from the simulation {\citep{SantosOlmsted2023}} showing routine bursts of star formation, while JD1B showed clear evidence of suppressed star formation with a relatively flat UV slope and weaker evidence for the presence of emission lines. The close projected distance of only 400$\,$pc in a halo that should be several kpc wide for stellar population of this mass might imply that the two regions formed in the same halo, but the SED fits SFHs with suppressed star formation for several dynamical times in only B. The difference in SFHs implies that both halos were subject to independent radiative and dynamical environments and likely formed much farther apart than 400 pc before coming closer or alternatively, both objects may be farther apart than their projected separation despite their coincident redshift. {Both scenarios, the SED model and the simulation} suggest an in-progress merger. This interaction merits further study with models tuned to investigate bursty star formation events and may yield more insight into high-redshift galaxy interactions and mergers.


\subsection{Prospects of future JWST observations}
The physical properties of MACS0647–JD inferred by our \JWST\ NIRCam observations indicate that future \JWST\ spectroscopy should allow the detection of several strong emission lines which would make it possible to improve constraints on the metallicity, gas ionization state, dust reddening, and star formation history of this intriguing $z\approx 11$ system. Based on the photometric redshift, the planned \JWST/NIRSpec prism observations extending to 5.3\um\ ($\sim$4300\AA\ rest-frame)
may detect strong emission lines like \CIIIw, \OIIw, H$\gamma$ and \NeIIIw+H$\zeta$+\HeIw\ (blended). 

Other strong rest-frame optical emission lines like \OIIIw\ and \Ha\ will fall in the wavelength domain of \JWST\ MIRI.
The inferred SFR of MACS0647–JD suggests that both \OIIIw\ and \Ha\ may be detected with MIRI medium-resolution spectroscopy (MRS) targeting JD1, which would then also allow simultaneous MIRI imaging of part of the MACS0647 cluster field. Such MIRI imaging could, depending on the position angle, also cover JD3, which is predicted to be sufficiently bright for detection in both F560W and F770W.

\section{Conclusions}
\label{sec:conclusions}

In this study, we report on public \JWST\ imaging observations 
of the $z \sim 11$ galaxy MACS0647--JD 
taken with 6 NIRCam filters (F115W, F150W, F200W, F277W, F356W, and F444W) by GO program 1433 (PI Coe).
Three lensed images are observed with magnifications \about~8, 5, and 2 and F356W AB mags of 25.1, 25.6, 26.6.
The delensed F356W magnitude is 27.3, with $M_{UV} = -20.4 \pm 17\%$.
MACS0647--JD is a J-band dropout appearing in all filters redward of F115W.
Its photometric redshift is $z = 10.6 \pm 0.3$ as estimated by 6 different methods.

MACS0647--JD is spatially resolved into two components A and B separated by \about~400 pc in projection.
They may be merging galaxies or two clumps that formed in situ within one galaxy.

Component A is brighter and very blue {($\beta \sim -2.6\pm0.1$)},
dust-free with a delensed radius \about~70{$^{+24}_{-24}\,$}pc and mass-weighted age \about~50\,Myr.
Component B is smaller and redder with perhaps some dust $A_V \sim 0.1$ mag,
a delensed radius \about~20{$^{+8}_{-5}\,$}pc, and mass-weighted age \about~100\,Myr.
Simulated galaxies with similar colors as observed for A and B {at similar redshift} have star formation histories {to be dissimilar} despite their proximity, {which is consistent with our SED fitting results},
suggesting they formed some distance apart, perhaps as separate galaxies observed now as they were on their way to merge.

Both have stellar masses \about$10^8\,M_\odot$ with A likely more massive by a factor of 2 or so.
With star formation rates on the order of 1\,$M_\odot$ yr\inv\ averaged over the past 10 Myr
and specific SFRs \about10 Gyr\inv, these galaxies are consistent with expectations
for the stellar main sequence at $z \sim 11$.
Given their small radii $< 100$ pc,
they have very high stellar mass surface densities, up to \about$10^4\,M_\odot$ pc$^{-2}$,
with correspondingly large SFR surface densities up to \about$~100\,M_\odot$ yr\inv\ kpc$^{-2}$.
These are large, though not exceeding theoretical limits or values measured for other extreme objects.

A small candidate companion galaxy C is identified \about~3 kpc away.
Three lensed images of C are at the expected locations are all J-band dropouts.
While fainter (F356W AB mag \about28) with more uncertain photometry,
its SED consistent with $z \sim 10.6$.

The NIRCam imaging spans 1--5\,\um\ to rest-frame 4300\,\AA\ at $z = 10.6$.
F444W is only partially redward of the Balmer break, 
limiting our ability to estimate ages and stellar masses.
Additional observations with the reddest NIRCam filter F480M and the NIRSpec MSA PRISM 
are upcoming and planned for January 2023.

\section{Acknowledgments}

We are grateful and indebted to all 20,000 people who worked to make \JWST\ an incredible discovery machine.

We dedicate these \JWST\ observations to Rob Hawkins, former lead developer of the Astronomer's Proposal Tool (APT). Rob lost his life in November 2020 while astronomers around the world were using APT to prepare observations we proposed for \JWST\ Cycle 1.

This work is based on observations made with the NASA/ESA/CSA 
\textit{James Webb Space Telescope} (\JWST)
and \textit{Hubble Space Telescope} (\HST). 
The data were obtained from the Mikulski Archive for Space Telescopes (MAST) 
at the Space Telescope Science Institute (STScI), 
which is operated by the Association of Universities for Research in Astronomy (AURA), Inc., 
under NASA contract NAS 5-03127 for \JWST. 
These observations are associated with program \JWST\ GO 1433
and \HST\ GO 9722, 10493, 10793, and 12101.

TH and A were funded by a grant for JWST-GO-01433 provided by STScI under NASA contract NAS5-03127.
LW acknowledges support from the National Science Foundation Graduate Research Fellowship under Grant No. DGE-2137419.
AA acknowledges support from the Swedish Research Council (Vetenskapsr\aa{}det project grants 2021-05559). PD acknowledges support from the NWO grant 016.VIDI.189.162 (``ODIN") and the European Commission's and University of Groningen's CO-FUND Rosalind Franklin program and warmly thanks the Institute for Advanced Study (IAS) Princeton, where a part of this work was carried out, for their generous hospitality and support through the Bershadsky Fund.  The Cosmic
Dawn Center is funded by the Danish National Research Foundation (DNRF) under grant \#140.  EZ and AV ackowledge support from the Swedish National Space Agency. MB acknowledges support from the Slovenian national research agency ARRS through grant N1-0238. MO acknowledges support from JSPS KAKENHI Grant Numbers JP22H01260, JP20H05856, JP20H00181, JP22K21349.
AZ, AKM and LJF acknowledge support by Grant No. 2020750 from the United States-Israel Binational Science Foundation (BSF) and Grant No. 2109066 from the United States National Science Foundation (NSF), and by the Ministry of Science \& Technology, Israel.
EV and MN acknowledge financial support through grants PRIN-MIUR 2017WSCC32, 2020SKSTHZ and INAF “main-stream” grants 1.05.01.86.20 and 1.05.01.86.31.
Y.J-T acknowledges financial support from the European Union's Horizon 2020 research and innovation programme 
under the Marie Sklodowska-Curie grant agreement No 898633, 
the MSCA IF Extensions Program of the Spanish National Research Council (CSIC),
and the State Agency for Research of the Spanish MCIU 
through the Center of Excellence Severo Ochoa award to the Instituto de Astrof\'isica de Andaluc\'ia (SEV-2017-0709). ACC thanks the Leverhulme Trust for their support via a Leverhulme Early Career Fellowship.


\facilities{JWST(NIRCam), HST(ACS, WFC3)}


\software{\astropy\ \citep{astropy2022, astropy2018, astropy2013},
          \photutils\ \citep{Bradley2022},
          \grizli\ \citep{Grizli},
          \eazypy\ \citep{Brammer2008},
          \piXedfit\ \citep{Abdurrouf2021,Abdurrouf2022}
          \bagpipes\ \citep{Carnall2018},
          \textsc{Cloudy}\citep{Ferland1998,Ferland2013,Ferland2017}, 
          Prospector \citep{Leja2017,Johnson2021},
          \beagle\ \citep{Chevallard2016_BEAGLE},
          \jdaviz\ \citep{JDAviz}
          }

\appendix

\section{Appendix}

\subsection{Photometry measurement for individual clumps}
\label{sec:no_use_photometry}
\begin{figure*}
\centering{
\includegraphics[width=0.75\textwidth]{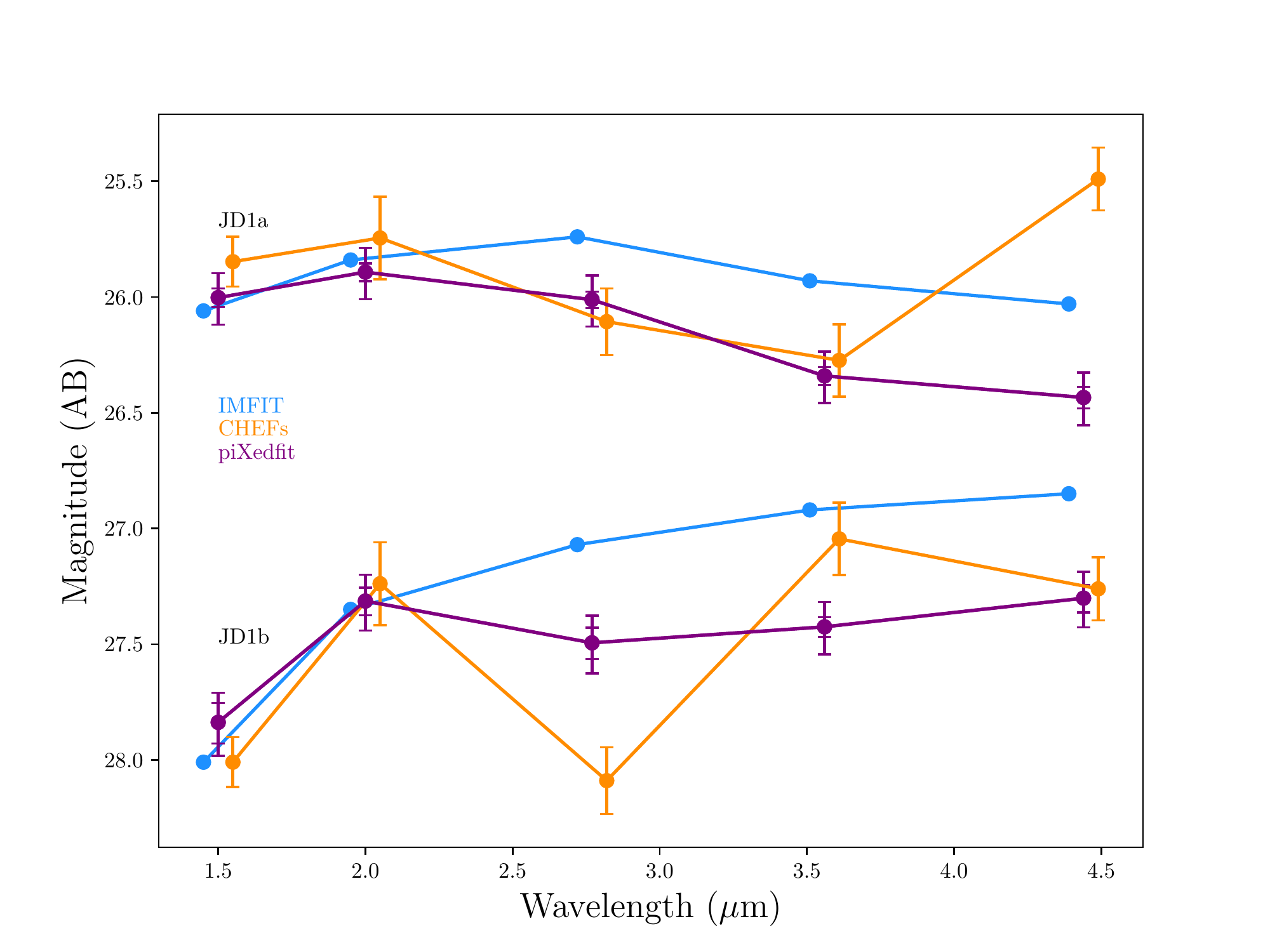}
}
\caption{
{Photometry of JD1A and JD1B measured from different methods, \piXedfit, IMFIT, and CHEFs. With 3\% and 10\% uncertainty added in quadrature are shown for \piXedfit.}
\label{fig:photometry_comparison}
}
\end{figure*}

{In Section. \ref{sec:photometry}, we measure the photometry for components A and B using different methods including \piXedfit, IMFIT, and CHEFs.
Only the photometry from \piXedfit\ was used for SED fitting to estimate physical properties.
Here we provide the comparison among three different methods, which is shown in Figure \ref{fig:photometry_comparison}.
We also test the SED fitting for the photometry from IMFIT and CHEFs.
The results are similar to the result using \piXedfit\ photometry.
}

\subsection{DELPHI Semi-Analytic Model}
\label{sec:DELPHI}

In brief, the DELPHI semi-analytic model \citep{Dayal2014,Dayal2022} uses a binary merger tree approach to 
build the dark matter assembly histories of $z \sim 4.5$ galaxies with halo masses ${\rm log}(M_h/ M_\odot)$ = 8 -- 14 up to $z \sim 40$. It then jointly tracks the build-up of dark matter halos and their baryonic components (gas, stellar, metal and dust mass) between $z \sim 40-4.5$ including the impact of both internal (supernova) and external (reionziation) feedback; here we consider a case that ignores reionization feedback since reionization affects $\lesssim20$\% of the volume of the Universe at $z \sim 10.5$ \citep{dayal2020}. The key strength of this model lies in its minimal free parameters (the star formation efficiency and fraction of supernova energy coupling to gas) and the fact that is it baselined against all available high-$z$ datasets including the evolving ultra-violet luminosity function, the stellar mass function and against the most recent dust estimates at $z \sim 7$ from the REBELS (Reionization Era Bright Emission Line Survey) ALMA large program \citep{bouwens2022}.

\subsection{GAINN Analog Simulated Galaxies}
\label{sec:hydrosims}

Synthetic fluxes from \JWST’s wideband filters F115W, F150W, F200W, F277W, F356W, and F444W were used as a training set for photometric redshift predictions of the combined fluxes of the A and B objects using the Galaxy Assembly and Interaction Neural Network (GAINN) method \citep{SantosOlmsted2023}, 
which trains deep convolutional neural networks on synthetic photometry created in post processing from in-situ star-forming ENZO \citep{Bryan2014} radiative-hydrodynamic simulations. GAINN was originally designed for redshifts higher than 11.4 and so star formation histories were shifted forward in time by 50, 100, and 150 Myr to produce a training set. Then a 20-layer network was trained on the approximately 12,000 SEDs in the set that fell between $z = 12$ and $z = 10$ to accommodate the likely redshift of MACS0647--JD, and achieved a mean absolute error of 0.0261 in the validation set. Predicted redshifts for  MACS0647--JD were consistent with galaxies at $z \sim 10.6$, returning a value of $z = 10.6417$ for object A and a value of $z = 10.5614$ for object B.

Additionally, $\chi^2$ SED-fitting to GAINN simulated galaxies was performed on JD1A and JD1B’s fluxes in F200W, F277W, F356W, and F444W, which focus on the rest-UV slope of both objects. Fitting was accomplished by first shifting over 10,000 synthetic observations in GAINN to $z=10.6$ and then using relative synthetic flux to create a mass and redshift-independent comparison. Then, the synthetic flux of the SED was scaled to the observed flux of JD1A and JD1B and $\chi^2$ values were calculated for the other three bands. We report four star formation histories corresponding to the lowest values of $\chi^2$ for both JD1A and JD1B along with their mass weighted mean stellar age after restricting our output to the best fit result in any particular halo tree branch. 

JD1B’s spectra was relatively rare in the simulation, resulting in various predictions with higher squared error, but it often matched to star formation histories with the strongest episodes of star formation ending earlier than 100 Myr before the observation, with some results showing a more recent burst. Predicted mean stellar ages were between 100 and 160 Myr, which is consistent with an absence of young stars and the flat UV slope of JD1B. 

All histories matched to JD1A featured bursty episodes of star formation peaking between 50 and 100 Myr before the observation, followed by a turn off and a resumption of star formation at the time of observation. Predictions for mean stellar age also converged between 45 and 80 Myr, which was consistent with the observed blue UV slope. All matched SED's had strong emission lines, implying that previous episodes of star formation well-enriched the ISM.

\bibliography{papers}{}
\bibliographystyle{aasjournal}


\end{document}

%% file: newcommands.tex
\newcommand{\LCDM}{$\Lambda$CDM}

\newcommand{\red}[1]{{\color{red} #1}}
\newcommand{\redss}[1]{{\color{red} ** #1}}
\newcommand{\redbf}[1]{{\color{red}\bf #1 \color{black}}}

\newcommand{\ny}{$\tilde {\rm n}$}
\newcommand{\about}{$\sim$}
\newcommand{\appr}{$\approx$}
\newcommand{\gt}{$>$}
\newcommand{\um}{$\mu$m}
\newcommand{\uJy}{$\mu$Jy}
\newcommand{\sig}{$\sigma$}
\newcommand{\Lya}{Lyman-$\alpha$}
\renewcommand{\th}{$^{\rm th}$}
\newcommand{\lam}{$\lambda$}

\newcommand{\tentothe}[1]{$10^{#1}$}
\newcommand{\tentotheminus}[1]{$10^{-#1}$}
\newcommand{\e}[1]{$\times 10^{#1}$}
\newcommand{\en}[1]{$\times 10^{-#1}$}

\newcommand{\sinv}{s$^{-1}$}

\newcommand{\tnm}[1]{\tablenotemark{#1}}
\newcommand{\super}[1]{$^{\rm #1}$}
\newcommand{\supa}{$^{\rm a}$}
\newcommand{\supb}{$^{\rm b}$}
\newcommand{\supc}{$^{\rm c}$}
\newcommand{\supd}{$^{\rm d}$}
\newcommand{\supe}{$^{\rm e}$}
\newcommand{\supf}{$^{\rm f}$}
\newcommand{\supg}{$^{\rm g}$}
\newcommand{\suph}{$^{\rm h}$}
\newcommand{\supi}{$^{\rm i}$}
\newcommand{\supj}{$^{\rm j}$}
\newcommand{\supk}{$^{\rm k}$}
\newcommand{\supl}{$^{\rm l}$}
\newcommand{\supm}{$^{\rm m}$}
\newcommand{\supn}{$^{\rm n}$}
\newcommand{\supo}{$^{\rm o}$}

\newcommand{\squared}{$^2$}
\newcommand{\cubed}{$^3$}

\newcommand{\sqarcmin}{arcmin\squared}

\newcommand{\supcomma}{$^{\rm ,}$}

\newcommand{\rhalf}{$r_{1/2}$}

\newcommand{\chisq}{$\chi^2$}

\newcommand{\per}{$^{-1}$}
\newcommand{\inv}{\per}
\newcommand{\Mstar}{$M^*$}
\newcommand{\Lstar}{$L^*$}
\newcommand{\phistar}{$\phi^*$}

\newcommand{\LUV}{$L_{UV}$}

\newcommand{\Msun}{$M_\odot$}
\newcommand{\Lsun}{$L_\odot$}
\newcommand{\Zsun}{$Z_\odot$}

\newcommand{\Mvir}{$M_{vir}$}
\newcommand{\Mt}{$M_{200}$}
\newcommand{\Mf}{$M_{500}$}

\newcommand{\Halpha}{H$\alpha$}
\newcommand{\Hbeta}{H$\beta$}
\newcommand{\Ha}{H$\alpha$}
\newcommand{\Hb}{H$\beta$}

\newcommand{\I}{\,{\sc i}}
\newcommand{\II}{\,{\sc ii}}
\newcommand{\III}{\,{\sc iii}}
\newcommand{\IV}{\,{\sc iv}}
\newcommand{\V}{\,{\sc v}}
\newcommand{\VI}{\,{\sc vi}}
\newcommand{\VII}{\,{\sc vii}}
\newcommand{\VIII}{\,{\sc viii}}

\newcommand{\HI}{H\,{\sc i}}
\newcommand{\HII}{H\,{\sc ii}}

\newcommand{\CII}{C\,{\sc ii}}
\newcommand{\CIII}{[C\,{\sc iii}]}
\newcommand{\CIIIw}{\CIII\,$\lambda$1908}
\newcommand{\CIV}{C\,{\sc iv}}
\newcommand{\OII}{[O\,{\sc ii}]}
\newcommand{\OIIw}{\OII\,$\lambda$3727}
\newcommand{\OIII}{[O\,{\sc iii}]}
\newcommand{\OIIIw}{\OIII\,$\lambda$5007}
\newcommand{\NeIII}{[Ne\,{\sc iii}]}
\newcommand{\NeIIIw}{\NeIII\,$\lambda$3869}
\newcommand{\HeIw}{HeI\,$\lambda$3889}


\newcommand{\Om}{\Omega_{\rm M}}
\newcommand{\OL}{\Omega_\Lambda}

\newcommand{\etal}{et al.}

\newcommand{\citeps}{\citep}

\newcommand{\HST}{{\em HST}}
\newcommand{\SST}{{\em SST}}
\newcommand{\Hubble}{{\em Hubble}}
\newcommand{\Spitzer}{{\em Spitzer}}
\newcommand{\Chandra}{{\em Chandra}}
\newcommand{\JWST}{{\em JWST}}
\newcommand{\Planck}{{\em Planck}}

\newcommand{\Bradac}{{Brada\v{c}}}

\newcommand{\citepeg}[1]{\citep[e.g.,][]{#1}}

\newcommand{\range}[2]{\! \left[ _{#1} ^{#2} \right] \!}  

\newcommand{\grizli}{\textsc{grizli}}
\newcommand{\eazypy}{\textsc{eazypy}}
\newcommand{\bagpipes}{\texttt{\scshape{Bagpipes}}}
\newcommand{\beagle}{\textsc{beagle}}
\newcommand{\photutils}{\textsc{photutils}}
\newcommand{\SEP}{\textsc{sep}}
\newcommand{\piXedfit}{\textsc{piXedfit}}
\newcommand{\astropy}{\textsc{astropy}}
\newcommand{\astrodrizzle}{\textsc{astrodrizzle}}
\newcommand{\multinest}{\textsc{multinest}}
\newcommand{\cloudy}{\textsc{Cloudy}}
\newcommand{\jdaviz}{\textsc{Jdaviz}}

\renewcommand{\tt}[1]{\texttt{#1}}

\newcommand{\SE}{\tt{SourceExtractor}}

\newcommand{\PD}[1]{\textcolor{blue}{[PD: #1\;]}}

%% file: observations.tex
\begin{deluxetable}{lllccr}
\tablecaption{\label{tab:obs}\HST\ and \JWST\ Exposure Times and Depths}
\tablewidth{\columnwidth}
\tablehead{
\colhead{}&
\colhead{}&
\colhead{$\lambda$}&
\colhead{Exp.}&
\colhead{Depth\tnm{a}}
\\[-6pt]
\colhead{Camera}&
\colhead{Filter}&
\colhead{(\um)}&
\colhead{Time (sec)}&
\colhead{(AB)}
}
\startdata
\HST\ WFC3/UVIS & F275W & 0.23--0.31  & 3879 & 27.4 \\
\HST\ WFC3/UVIS & F336W & 0.30--0.37  & 2498 & 27.6 \\
\HST\ WFC3/UVIS & F390W & 0.33--0.45  & 2545 & 28.1 \\
\HST\ ACS/WFC & F435W & 0.36--0.49 & 2124 & 28.0 \\
\HST\ ACS/WFC & F475W & 0.39--0.56  & 2248 & 28.2 \\
\HST\ ACS/WFC & F555W & 0.46--0.62  & 7740 & 28.7 \\
\HST\ ACS/WFC & F606W & 0.46--0.72  & 2064 & 28.3 \\
\HST\ ACS/WFC & F625W & 0.54--0.71  & 2131 & 27.9 \\
\HST\ ACS/WFC & F775W & 0.68--0.86  & 2162 & 27.8 \\
\HST\ ACS/WFC & F814W\tnm{b} & 0.69--0.96  & 8800 & 28.5 \\
\HST\ ACS/WFC & F850LP & 0.80--1.09  & 4325 & 27.3 \\
\HST\ WFC3/IR & F105W & 0.89--1.21   & 2914 & 28.3 \\
\HST\ WFC3/IR & F110W & 0.88--1.41   & 1606 & 28.7 \\
\HST\ WFC3/IR & F125W & 1.08--1.41   & 2614 & 28.3 \\
\HST\ WFC3/IR & F140W & 1.19--1.61   & 2411 & 28.7 \\
\HST\ WFC3/IR & F160W & 1.39--1.70   & 5229 & 28.4 \\
\JWST\ NIRCam & F115W & 1.0--1.3   & ~\,2104 & 28.1 \\
\JWST\ NIRCam & F150W & 1.3--1.7   & ~\,2104 & 28.3 \\
\JWST\ NIRCam & F200W & 1.7--2.2   & ~\,2104 & 28.4 \\
\JWST\ NIRCam & F277W & 2.4--3.1   & ~\,2104 & 28.9 \\
\JWST\ NIRCam & F356W & 3.1--4.0   & ~\,2104 & 29.0 \\
\JWST\ NIRCam & F444W & 3.8--5.0   & ~\,2104 & 28.8
 \enddata
\tablenotetext{a}{5\sig\ point source AB magnitude limit (within a 0.2\arcsec\ diameter aperture).}
\tablenotetext{b}{We excluded one dataset (J8QU04020:~3960\,s) because it is contaminated by scattered light from the WFPC2 internal lamp, which was in use for a parallel program.}
\end{deluxetable}

%% file: recalibration.tex
\begin{deluxetable}{lcc}
\tablecaption{\label{tab:recalibration}Photometric Recalibration}
\tablewidth{\columnwidth}
\tablehead{
\colhead{Filter}&
\colhead{JD1, JD2}&
\colhead{JD3}
}
\startdata
F115W & A3 0.9687 & A1 0.9826 \\
F150W & A3 0.9536 & A1 0.9777 \\
F200W & A3 0.9658 & A1 0.9891 \\
F277W & A5 1.0239 & A5 1.0239 \\
F356W & A5 0.9763 & A5 0.9763 \\
F444W & A5 1.0073 & A5 1.0073
 \enddata
\tablecomments{We multiply JD1,2,3 fluxes and uncertainties
by these values to correct from \grizli\ v4 calibration
to \texttt{jwst\_0995.pmap}.}
\end{deluxetable}

%% file: photometry.tex
\begin{deluxetable}{lllccr}
\tablecaption{\label{tab:pho}Coordinates and multiwavelength photometry of MACS0647--JD}
\tablewidth{\columnwidth}
\tablehead{
\colhead{Filter}&
\colhead{JD1}&
\colhead{JD2}&
\colhead{JD3}&
}
\startdata
RA(J2000)  & 101.9822676 &  101.971326 & 101.9811153 &  \\
DEC(J2000) & 70.24328239 &  70.2397157 & 70.26059029 &  \\
F275W & $-31\pm35$ &  $11\pm30$& $43\pm27$ & \\
F336W & $-8\pm27$ &  $32\pm24$& $-24\pm21$ & \\
F390W &$-12\pm17$ &  $-5\pm44$& $24\pm23$ &\\
F435W& $-25\pm21$ &  $70\pm18$& $-23\pm21$ &   \\
F475W & $7\pm15$ &  $-26\pm113$& $21\pm12$ &  \\
F555W & $-2\pm10$ &  $15\pm8$& $8\pm7$ &  \\
F606W & $-16\pm10$ &  $4\pm12$& $-1\pm28$ &  \\
F625W & $-38\pm20$ &  $-24\pm17$& $15\pm17$ &\\
F775W & $24\pm28$ &  $-16\pm21$ & $-18\pm17$ &  \\
F814W & $-8\pm9$ &  $0\pm8$& $-3\pm7$ &   \\
F850LP & $-55\pm33$ &  $-11\pm30$& $29\pm30$ &\\
F105W &  $3\pm19$ &  $4\pm16$& $-1\pm14$ &  \\
F110W & $-25\pm13$ &  $5\pm12$& $7\pm10$ &  \\
F115W & $-1\pm8$ &  $-35\pm7$& $-2\pm6$ &  \\
F125W& $-13\pm19$ &  $-9\pm17$& $6\pm14$ &  \\
F140W &$126\pm11$ &  $97\pm12$& $59\pm9$& \\
F150W &$304\pm7$ &  $173\pm6$& $92\pm5$ &  \\
F160W & $301\pm17$ &  $203\pm17$& $85\pm13$&  \\
F200W & $354\pm6$ &  $227\pm6$& $115\pm4$ & \\
F277W &$351\pm6$ &  $248\pm5$ &$93\pm3$ &\\
F356W & $322\pm6$ &  $203\pm5$& $86\pm3$ &\\
F444W& $330\pm9$ &  $214\pm7$& $95\pm5$ &
     \enddata
\tablenotetext{}{{\bf Note.} All fluxes are given in $\rm nJy$.}
\end{deluxetable}

%% file: photometry_individual.tex
\begin{deluxetable*}{lllcccr}
\tablecaption{\label{tab:phoindi}Multiwavelength photometry of two individual galaxies in MACS0647--JD (i.e., JDa and JDb). }
\tablewidth{\columnwidth}
\tablehead{
\colhead{Filter}&
\colhead{JD1a}&
\colhead{JD1b}&
\colhead{JD2a}&
\colhead{JD2b}&
\colhead{JD3a}&
\colhead{JD3b}
}
\startdata
F435W& $-16\pm6$ &  $-2\pm5$ & $4\pm5$ &  $8\pm5$ & $-5\pm4$ & $-3\pm4$  \\
F475W & $-9\pm5$ &  $-2\pm3$ & $-10\pm4$ & $-9\pm3$ & $-3\pm3$ & $0\pm2$ \\
F555W & $-10\pm3$ & $-5\pm2$ & $-3\pm3$ & $-5\pm2$ & $-1\pm2$ & $-2\pm1$ \\
F606W & $-6\pm7$ &  $-2\pm4$ & $0\pm4$ & $-1\pm3$ & $6\pm14$ & $1\pm3$ \\
F625W & $-19\pm7$ &  $-4\pm4$ & $2\pm5$ & $-6\pm4$ & $8\pm5$ & $-3\pm2$\\
F775W & $-23\pm9$ &  $-6\pm8$ & $-25\pm6$ & $3\pm5$ & $-8\pm5$ & $-3\pm3$ \\
F814W & $-15\pm4$ &  $-5\pm3$ & $-8\pm3$ & $-4\pm2$ & $-1\pm2$ &  $-1\pm2$ \\
F115W & $-18\pm3$ &  $-2\pm2$ & $-10\pm3$ & $-7\pm2$ & $-2\pm2$ & $-3\pm1$ \\
F150W & $144\pm3$ &  $27\pm2$ & $91\pm3$ & $16\pm2$ & $55\pm2$ & $7\pm1$ \\
F200W & $160\pm3$ &  $43\pm2$ & $115\pm2$ & $27\pm2$ & $60\pm2$ & $13\pm1$ \\
F277W & $143\pm2$ &  $37\pm2$ & $90\pm2$ & $27\pm1$ & $43\pm1$ & $8\pm1$\\
F356W & $106\pm2$ &  $39\pm1$ & $64\pm2$ & $25\pm1$ & $30\pm1$ & $9\pm1$\\
F444W & $97\pm3$ &  $44\pm2$ & $59\pm2$ & $26\pm1$ & $28\pm1$ & $8\pm1$
 \enddata
\tablenotetext{}{{\bf Note.} All fluxes are described in $\rm nJy$
and were measured using piXedfit \citep{Abdurrouf2021}.
}
\end{deluxetable*}

%% file: sedfitting.tex
\begin{deluxetable*}{lccccc}
\tablecaption{\label{tab:sedfitting}SED fitting methods, templates, and assumptions used}
\tablewidth{\columnwidth}
\tablehead{
\colhead{Method}&
\colhead{SEDs}&
\colhead{IMF}&
\colhead{SFH}&
\colhead{Dust}&
\colhead{Ionization log($U$)}
}
\startdata
EAZY & SFHZ & \\
\bagpipes\ & BPASS+\textsc{Cloudy} & \citet{1993MNRAS.262..545K} & delayed tau & \citet{Salim18} & $-4$ to $-1$\\
\piXedfit\ & FSPS+\textsc{Cloudy} & \citet{Chabrier2003} & double power-law & \citet{Charlot2000} & $-2$ \\
Prospector & FSPS+\textsc{Cloudy} & \citet{Chabrier2003} & constant / non-parametric & SMC \citep{Pei1992} & $-4$ to $-1$ \\
BEAGLE & BC03+\textsc{Cloudy} & \citet{Chabrier2003} & constant & SMC \citep{Pei1992} & $-4$ to $-1$ \\
\enddata
\end{deluxetable*}

%% file: properties.tex
\begin{deluxetable*}{lllccr}
\tablecaption{\label{tab:phy}Physical properties of MACS0647--JD estimated from \texttt{Bagpipes} corrected for magnification (flux ratio). }
\tablewidth{\columnwidth}
\tablehead{
\colhead{}&
\colhead{JD1}&
\colhead{JD2}&
\colhead{JD3}&
\colhead{Combined}&
}
\startdata
Formation age ($\rm Myr$) &$160^{+69}_{-69}$& $151^{+60}_{-74}$ & $153^{+121}_{-75}$& $177^{+73}_{-71}$  \\
Mass-weighted age ($\rm Myr$) & $54^{+24}_{-23}$& $51^{+20}_{-25}$ & $51^{+43}_{-25}$& $59^{+25}_{-24}$  \\
Stellar Mass (${\rm log}(M_{*}/M_{\odot})$) & $8.62^{+0.11}_{-0.15}$ & $8.61^{+0.10}_{-0.18}$ &  $8.51^{+0.18}_{-0.22}$& $8.65^{+0.11}_{-0.14}$    \\
SFR ($M_\odot/{\rm yr}$) within 100 Myr &$4^{+1}_{-1}$& $4^{+1}_{-1}$ &  $3^{+1}_{-1}$&  $4^{+1}_{-1}$  \\
log sSFR (${\rm yr}^{-1}$) &$-7.98^{+0.05}_{-0.09}$ &$-7.96^{+0.04}_{-0.08}$ &  $-7.97^{+0.04}_{-0.16}$& $-8.00^{+0.07}_{-0.10}$ \\
Photometric Redshift &$10.64^{+0.06}_{-0.07}$ &$10.76^{+0.08}_{-0.09}$ & $10.50^{+0.14}_{-0.16}$  & $10.65^{+0.09}_{-0.10}$   \\
Relative flux ($\ge \rm F200W$) & 1 & 0.66 & 0.28 & 1.94\\
Magnification (flux ratio) & 8 & 5.3 & 2.2& 15.5 \\ 
{Magnification (lens model)} & 6.9 & 6.3 & 2.1 & 15.3 \\ 
{Tangential Magnification} & 4.7 & 4.4 & 1.8 \\
\enddata
\tablenotetext{}{
{{\textbf{Note}. 
Magnification uncertainties are on the order of 15\% \citep{Meneghetti2017}
and are not included in the uncertainties quoted above for stellar mass and SFR.
}}}
\end{deluxetable*}

%% file: individual.tex
\begin{deluxetable*}{lclcccccc}
\tablecaption{\label{tab:indi_phy}Physical properties of A and B analyzed individually by various methods
assuming $z = 10.6$.
Photometry is measured in the brightest image JD1 by \piXedfit,
analyzed with and without inflated uncertainties and corrected for magnification.}
\tablewidth{\columnwidth}
\tablehead{
\colhead{} &
\colhead{Photometric} &
\colhead{} &
\colhead{} &
\colhead{Age$^a$} &
\colhead{Stellar Mass} &
\colhead{specific SFR$^b$} &
\colhead{Dust}
\vspace{-0.07in}\\
\colhead{Clump} &
\colhead{Uncertainty} &
\colhead{Method} &
\colhead{Star Formation History} &
\colhead{Myr} &
\colhead{log($M_{*}/M_\odot$)} &
\colhead{Gyr\inv} &
\colhead{$A_{V}$ mag} &
\colhead{$\chi^{2}$}
}
\startdata
JD1A & \nodata & \bagpipes  & delayed $\tau$ exponential & $1.3 \pm 0.2$            & $7.2 \pm 0.1$       & $101^{+1}_{-1}$ &  $0.00^{+0.01}_{-0.00}$ & $93$\\
     &  +3\%   & \bagpipes  & delayed $\tau$ exponential & $1^{+1}_{-1}$     & $7.2^{+0.1}_{-0.1}$ & $101^{+1}_{-1}$ &  $0.00^{+0.01}_{-0.00}$ & $82$ \\
     & +10\%   & \bagpipes  & delayed $\tau$ exponential & $14^{+14}_{-9}$      & $7.7^{+0.2}_{-0.3}$  &$48^{+52}_{-22}$  &  $0.01^{+0.02}_{-0.01}$ & $67$\\
     & \nodata & \piXedfit  & double power-law           & $23^{+19}_{-9}$      & $8.2^{+0.1}_{-0.1}$ & $8^{+9}_{-7}$ &  $0.34^{+0.22}_{-0.21}$ & $105$ \\
     & \nodata & Prospector & non-parametric cont.       & $43^{+25}_{-20}$     & $8.3 \pm 0.1$       & $0$ &  $0.01^{+0.02}_{-0.01}$ & $18^{c}$\\
     & \nodata & Prospector & constant                   & $35^{+8}_{-6}$       & $8.0 \pm 0.1$       & $14\pm3$ &  $0.00^{+0.00}_{-0.00}$ & $90^{c}$ \\
     & \nodata & \beagle    & constant                   & $35 \pm 7$           & $8.1 \pm 0.1$       &$14^{+4}_{-3}$      &  $0.00^{+0.00}_{-0.00}$ & $103^{c}$\\
JD1B & \nodata & \bagpipes   & delayed $\tau$ exponential & $70^{+49}_{-44}$    & $7.8^{+0.1}_{-0.3}$       & $11^{+16}_{-4}$  &  $0.10^{+0.06}_{-0.03}$ & $15$ \\
     & +3\%    & \bagpipes   & delayed $\tau$ exponential & $67^{+45}_{-40}$    & $7.8^{+0.2}_{-0.3}$ & $12^{+16}_{-5}$ &  $0.10^{+0.07}_{-0.04}$ &$14$\\
     & +10\%   & \bagpipes   & delayed $\tau$ exponential & $81^{+41}_{-50}$    & $7.8^{+0.1}_{-0.3}$       & $10^{+14}_{-3}$ &  $0.10^{+0.08}_{-0.05}$ & $12$ \\
     & \nodata & \piXedfit   & double power-law           & $111^{+32}_{-30}$   & $8.3^{+0.1}_{-0.3}$ &$0^{+2}_{-0}$ &  $0.25^{+0.14}_{-0.15}$ & $20$\\
     & \nodata & Prospector  & non-parametric cont.       & $127^{+50}_{-58}$   & $8.1 \pm 0.2$       & $3^{+10}_{-2}$ &  $0.12^{+0.12}_{-0.08}$ & $2.9^{c}$\\
     & \nodata & Prospector  & constant                   & $3^{+35}_{-2}$      & $6.9 \pm 0.2$       &$185^{+319}_{-171}$ &  $0.03^{+0.03}_{-0.02}$ & $1.5^{c}$\\
     & \nodata & \beagle     & constant                   & $10^{+49}_{-8}$     & $7.3 \pm 0.6$       & $51^{+198}_{-42}$ &  $0.01^{+0.03}_{-0.01}$ &$0.6^{c}$
\enddata
\tablenotetext{a}{Mass-weighted.}
\tablenotetext{b}{Within recent 10\,Myr.}
\tablenotetext{c}{Calculated in F200W and redder photometry.}
\textbf{Note.} 
Most of these SED fitting methods assume a \citet{Chabrier2003} IMF (Table \ref{tab:sedfitting}).
BAGPIPES assumes \citet{1993MNRAS.262..545K};
to renormalize those results, we multiplied the stellar masses by 0.94
\citep{Madau2014}. 

\end{deluxetable*}

%% file: clumps.tex
\begin{deluxetable*}{lccrcr}
\tablecaption{\label{tab:clumps}Estimated clump properties adopting fiducial stellar mass and star formation rate
}
\tablewidth{\columnwidth}
\tablehead{
\colhead{Clump} &
\colhead{Radius} &
\colhead{Stellar Mass} &
\colhead{Stellar Mass Density} &
\colhead{SFR} &
\colhead{SFR Density}
}
\startdata
JD1A 
& 70{$^{+24}_{-24}\,$}pc 
& $10^8 M_\odot$ 
& 1800 $M_\odot$ pc$^{-2}$
& 1 $M_\odot$ yr$^{-1}$ 
& 18 $M_\odot$ yr$^{-1}$ kpc$^{-2}$ \\
JD1B 
& 20{$^{+8}_{-5}\,$}pc 
& $6\times 10^7 M_\odot$ 
& 12000 $M_\odot$ pc$^{-2}$
& 0.6 $M_\odot$ yr$^{-1}$ 
& 120 $M_\odot$ yr$^{-1}$ kpc$^{-2}$ \\
\enddata
\tablenotetext{}{
{{\textbf{Note}. 
The estimated stellar mass and SFR vary from different SED fitting packages under different assumptions (e.g., star formation history).
The uncertainties of stellar mass density and SFR density may be up to 200\%.
}}}
\end{deluxetable*}